\documentclass{article}

\usepackage[natbib=true,sorting=none,style=numeric,mincitenames=1,maxcitenames=2]{biblatex}
\DeclareAutoCiteCommand{inline}[r]{\parencite}{\parencites}

\usepackage{xcolor,latexsym,epsfig,amssymb,amsmath,color,subcaption,pictex,amsfonts,multicol}
\usepackage{upgreek}
\usepackage[inline]{enumitem}
\usepackage{amsfonts}
\usepackage{amsthm}
\usepackage{amscd}
\usepackage{ulem}
\usepackage{cancel}
\usepackage{amssymb}
\usepackage{graphicx}
\usepackage{epsfig}
\usepackage{graphics}
\usepackage{setspace}
\usepackage{anyfontsize}
\usepackage{mathrsfs}

\usepackage[utf8]{inputenc} 
\usepackage{amsmath} 
\usepackage{amssymb} 
\usepackage{microtype} 
\usepackage{graphicx} 
\usepackage[a4paper, margin=2.5cm]{geometry}
\usepackage{arxiv} 

\newcommand{\dst} {\displaystyle}

\bibliography{references.bib}

\newcommand{\ie}{{\it i.e.}\ }

\newcommand{\be}{\begin{eqnarray}}
\newcommand{\ee}{\end{eqnarray}}

\newcommand{\bes}{\begin{eqnarray*}}
\newcommand{\ees}{\end{eqnarray*}}

\newcommand{\non} \nonumber

\newcommand{\tp}{t_+^0}
\newcommand{\De}{D_{\rm{e}}}

\newcommand{\mrs}{\mathrm{s}}
\newcommand{\mre}{\mathrm{e}}
\newcommand{\mrn}{\mathrm{n}}
\newcommand{\mrp}{\mathrm{p}}

\newcommand{\tdf}{1+\frac{\partial\ln f_{\pm}}{\partial\ln c_\mre}}

\DeclareUnicodeCharacter{2212}{-}

\begin{document}
\title{Parameterising continuum level Li-ion battery models \& the {\tt LiionDB} database}
\author{A.~A. Wang, S.~E.~J. O'Kane, F. Brosa Planella, J. Le Houx, K. O'Regan, M. Zyskin,\\ J. Edge, C.~W. Monroe, S.~J. Cooper, D.~A. Howey, E. Kendrick, J.~M. Foster}
\date{\today}
\maketitle

\begin{abstract}
The Doyle--Fuller--Newman framework is the most popular physics-based continuum-level description of the chemical and dynamical internal processes within operating lithium-ion-battery cells. With sufficient flexibility to model a wide range of battery designs and chemistries, the framework provides an effective balance between detail, needed to capture key microscopic mechanisms, and simplicity, needed to solve the governing equations at a relatively modest computational expense. Nevertheless, implementation requires values of numerous model parameters, whose ranges of applicability, estimation, and validation pose challenges. This article provides a critical review of the methods to measure or infer parameters for use within the isothermal DFN framework, discusses their advantages or disadvantages, and clarifies limitations attached to their practical application. Accompanying this discussion we provide a searchable database, available at  {\tt www.liiondb.com}, which aggregates many parameters and state functions for the standard Doyle--Fuller--Newman model that have been reported in the literature.
\end{abstract}

\section{Introduction}

Lithium-ion batteries (LiBs) are becoming increasingly important to address the climate emergency. Their prominence will grow over the coming decades, a time during which internal combustion engine sales will likely be discouraged across large parts of the world, causing electric-vehicle uptake to expand, and batteries integrated into power grids and off-grid systems will increasingly support renewable electricity generation. The growing importance of LiB technology motivates a search for methods to improve device design, with the aims of increasing power density, energy density, temperature range and longevity. As well as motivating changes in the chemistry of active materials within electrodes, design improvements require the careful optimisation of many device and system characteristics. At the cell level, optimisation might consider tab placement; cathode/anode pairing and balancing; electrode thickness, geometry, porosity, and active-material loading; separator thickness, porosity, and microstructure; and quantity or composition of electrolyte \autocite{ue2020basic}. At the system level, performance optimisation must not only consider the size, form factor and number of cells and their electrical arrangement, but also the overall thermal-management strategy and the methods used for performance monitoring and control. Identifying how these characteristics interact to determine the overall behaviour of a battery in a given application poses a formidable problem. Whilst this can be approached from a purely experimental standpoint, essentially by trial and error, significantly accelerated design is enabled by predictive simulations that facilitate testing in a virtual environment, bypassing the need for slow and costly physical prototyping.

Modelling can rationalize and predict many aspects of battery performance and control, as well as improving the interpretation of experimental results and making possible the testing of novel designs \textit{in silico}. For example, models can explore the operational boundaries of a new design or show how temperature changes might impact performance. They may be used to predict lifetime, estimate state of charge (SoC) 
or power capability, and calculate returns on investment. There are many questions one might ask of models, for example, `what is the energy efficiency when this battery is used in a particular way?', `how do cell-design choices impact the overall rate capability of this battery chemistry?', and `what is the expected lifetime of this cell?'. Even physics-based models are not a panacea, however: they require careful structuring and parameterisation specific to the device being considered, its expected operating regimes, and the modelling objective itself.

Theoretical approaches relevant to LiBs have been developed across a range of length scales, spanning from the atomic level up to the pack level. 
For example, at the atomistic scale, electronic structure calculations based on density functional theory and Monte Carlo methods are commonly used to compute intrinsic material properties, whereas packs are usually described by equivalent circuit models (ECMs). Intermediate to these two extremes is the cell level, i.e., the scale of electrode pairs. Modelling at this scale will be the primary focus of this review. 

Various models exist to describe the cell level, each suitable for answering a different sort of research question. If the application is relatively benign (low power with no ambient-temperature extremes), then a simple model might suffice. For example, if a use case requires low  current densities and a constant load, then voltage response is likely to be dominated by the open-circuit voltage (OCV) and the cell's internal impedance; a more complex dynamical model, with outputs dependent on the instantaneous SoC or the current-load history, is unnecessary. When considering more demanding applications in motorsports or aerospace, or for long-term-performance prediction, the assumptions made within simpler models can become invalid, however. More finely grained physical descriptions at the level of individual materials within the battery may be needed to achieve sufficient predictive accuracy. The task of the battery modeller is to consider the research questions being asked, select an appropriate model structure to address these questions parsimoniously, identify which parameters within the structure are key, and discover --- via experiment, literature, or fitting --- the parameter values.

The most ubiquitous cell-level modelling framework for LiBs is the Doyle--Fuller--Newman (DFN) model, called `Dualfoil' by its authors, but typically named `the Newman model' in the wider literature, or sometimes the `pseudo-two-dimensional' (P2D) model, owing to its consideration of physics on both macroscopic and microscopic scales. Here we shall use the DFN acronym. The DFN framework derives from porous-electrode theory, stemming from John Newman's 1961 Master's thesis under Charles Tobias \autocite{Newman1961,newman1962}, with subsequent development specific to LiBs being laid out in four foundational works during the 1990s \autocite{doyle1993modeling,fuller1994simulation,Ma_1995,Doyle1995}. Rigorous mathematical justification is also provided by recent works \autocite{richmulti,ciucci,Daz2018}. The DFN framework strikes an excellent balance between speed and complexity, incorporating detailed physics beyond ECMs but --- unlike density-functional-theory or phase-field simulations --- remaining coarse enough for efficient device-level simulations \autocite{GLi2020}. 

The basic isothermal DFN model (the focus of this review) comprises a system of coupled partial differential equations (PDEs) and can contain upwards of 20 scalar parameters and around five scalar state functions. Complete parameterisation is difficult, as many requisite quantities must be inferred indirectly from experimental voltage and current data. Moreover, there can be significant device-to-device variability, even in cells that are ostensibly prepared in the same way. Thus, proper parameterisation is a hurdle that must be addressed to realise the maximum benefits from DFN models.


Many in the battery modelling community tend to rely on the scientific literature to find parameters, rather than carrying out their own experiments. Locating the required values in a rapidly growing body of work can be time-consuming. More importantly, great care must be taken to ensure that parameter values are extracted from raw data under consistent assumptions, to avoid undermining the validity of a model that incorporates them. 

The primary aim of this review is to alleviate the DFN parameterisation bottleneck in two ways. First, we survey the experimental techniques employed for parameterisation, clarifying the often unstated assumptions embedded within them. This allows a better understanding of their theoretical basis and empowers readers to decide what measurements can be reliably applied to which operating regimes. Second, we provide a searchable database that tabulates many parameter measurements available from the literature. The interactive database is populated by an extensive catalogue of reported parameters, as well as the details of measurement techniques used and materials characterised. Future implementations will allow the information source to continue to grow as LiB research and DFN modelling evolves.

The review proceeds with a brief outline of the isothermal DFN framework. Three subsequent sections focus on geometric (section \ref{sec:geomparams}), electrode (section \ref{sec:electrodeparams}) and electrolyte (section \ref{sec:elyteparams}) parameters. We discuss the definition and significance of each parameter and survey reported values, providing discussion about literature consensus and any inherent variability. We then dissect the common experimental measurement approaches and critique their applicability and reliability. Finally, in section \ref{bass}, we present the database ({\tt www.liiondb.com} \autocite{LiionDB}) that was developed as part of this work, before drawing our conclusions.

\section{\label{PEToutline}The DFN model}

\begin{figure}
  \centering
  \includegraphics[width=1\textwidth]{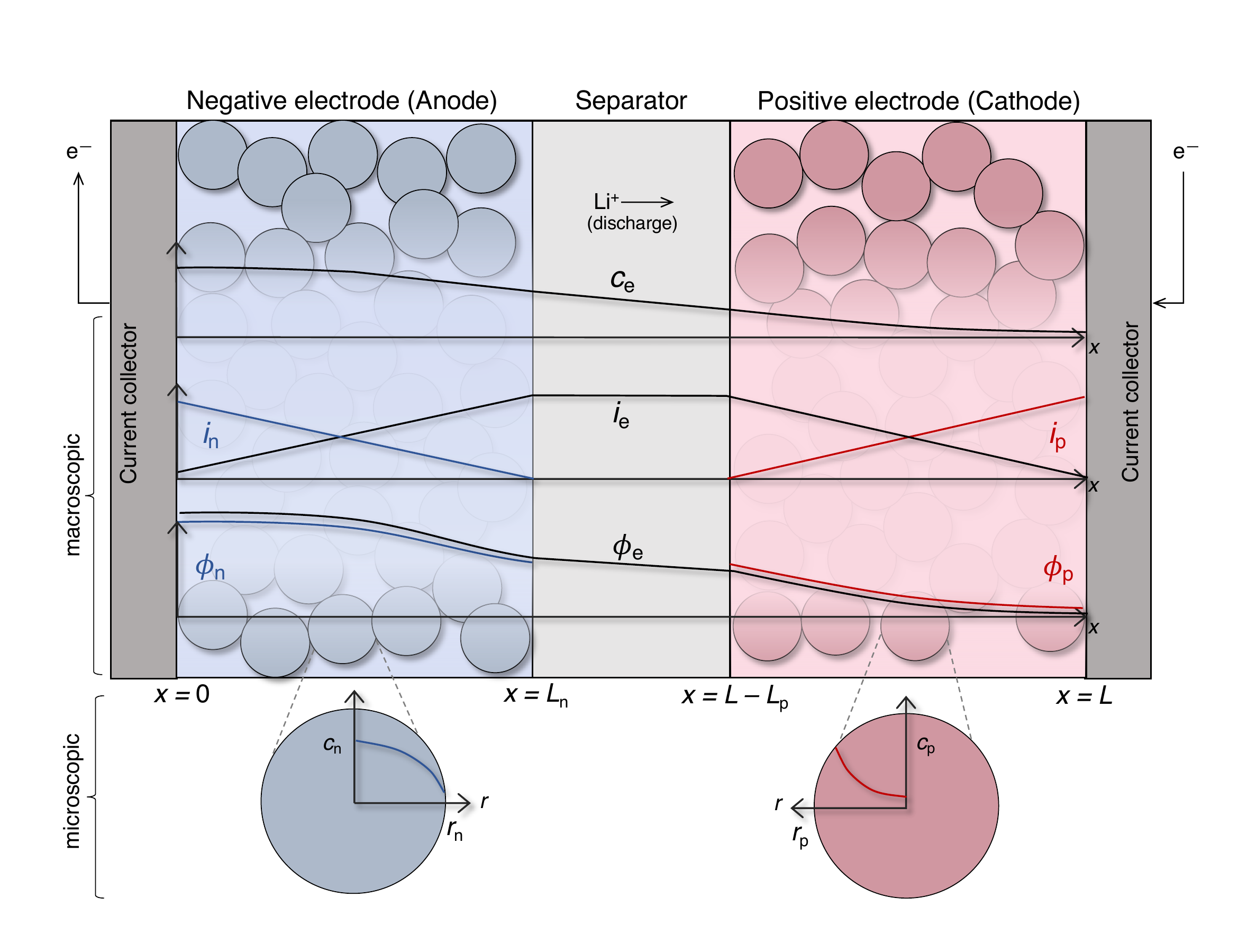}  
\caption{Schematic of Doyle-Fuller-Newman (DFN) cell geometry, indicating the macroscopic and microscopic coordinate systems and domains of definition for the different variables. The positions $x=0$, $x=L_\mrn$, $x=L-L_\mrp$, and $x=L$ represent domain interfaces between current collector and anode, anode and separator, separator and cathode and cathode and current collector, respectively. Plotted above the particles are position-dependent variables during a discharge: Li$^+$ concentration $c$, current density $i$ and potential distribution $\phi$. The subscripts $\mre$, $\mrn$ and $\mrp$ denote the electrolyte, negative electrode and positive electrode, respectively.}
\label{fig:geom}
\end{figure}

Variables and parameters involved in the standard DFN model are summarised in Tables \ref{table1} and \ref{table2}, respectively; the various material domains described within the framework are listed in Table \ref{subscripts}. Figure \ref{fig:geom} illustrates the domains in a cell schematically, with porous negative-electrode (an anode during discharge), separator and positive-electrode (cathode) regions spanning positions between points $x=0$, $x=L_\mrn$, $x=L-L_\mrp$ and $x=L$. 

\begin{table}
\centering
\caption{Description of the variables used in the formulation of the full cell DFN model.}
\label{table1}
\begin{tabular}{|c|c|c|}
\hline \textbf{Variable}  & \textbf{Description}  & \textbf{Units} \\
\hline $c_k(x,r,t)$ & Li$^+$ Concentration in electrode particles & mol/m$^3$\\
\hline $c_{k\mathrm{s}(x,t)}$ & Concentration at electrode particle surface & mol/m$^3$\\
\hline $c_\mre(x,t)$ & Li$^+$ concentration in electrolyte & mol/m$^3$\\
\hline $N_\mre(x,t)$ & Molar flux of ions in electrolyte & mol/m$^2$/s\\
\hline $i_k(x,t)$ & Current density in electrodes & A/m$^2$\\ 
\hline $i_\mre(x,t)$ & Current density in electrolyte & A/m$^2$\\
\hline $j_k(x,t)$ & Reaction current density & A/m$^2$\\
\hline $j_{k0}(x,t)$ & Exchange current density & A/m$^2$\\
\hline $\phi_k(x,t)$ & Electrode potential & V\\
\hline $\phi_\mre(x,t)$ &  Electrolyte potential & V\\
\hline $T(x,t)$ & Temperature & K\\
\hline $\eta_k(x,t)$ & Overpotential at electrode-electrolyte interface & V\\
\hline
\end{tabular}
\end{table}

\begin{table}
\centering
\caption{Functions and parameters for full cell DFN model}
\label{table2}
\begin{tabular}{|c|c|c|}
\hline \textbf{Param./}  & \textbf{Description}  & \textbf{Value and Units} \\
 \textbf{Ftn.}  &  &  \\
\hline $F$ & Faraday's constant & $9.6485\times 10^4 $(A\,s\,mol$^{-1}$)\\
\hline $R$ & Universal gas constant & $8.3145 $ (J\,K$^{-1}$mol$^{-1}$)  \\
\hline $A$ & Planar electrode area & m$^2$\\
\hline $b_k$ & Particle surface area per unit volume & 1/m\\
\hline $c^{\max}_{k}$ & Maximum Li$^+$ concentration in particle & mol/m$^3$\\
\hline $c_{\mre 0}$ & Initial/rest Li$^+$ concentration in electrolyte & mol/m$^3$\\
\hline $c_{k 0}$ & Initial Li$^+$ concentration in electrode & mol/m$^3$\\
\hline $D_k$ & Li$^+$ diffusivity in particle & m$^2$/s\\
\hline $D_\mre$ & Li$^+$ diffusivity in electrolyte & m$^2$/s\\
\hline $i_\mathrm{app}$ & Applied current density & A/m$^2$\\
\hline $K_k$ & Reaction rate constant & mol m$^{-2}$ s$^{-1}$\\
\hline $L_k$ & Electrode and separator thicknesses & m\\
\hline $L$ & Total cell thickness ($L_\mrn+L_\mrs+L_\mrp$) & m\\
\hline $r_k$ & Particle radius & m\\
\hline $\tp$ & Li$^+$ transference number & -\\
\hline $U_k$ & Open-circuit potential (OCP) & V\\
\hline $\mathcal{B}_k$ & Liquid phase transport efficiency& -\\
\hline $\varepsilon_k$ & Electrolyte volume fraction (porosity) & -\\
\hline $\varepsilon_{\textrm{act},k}$ & Active material volume fraction & -\\
\hline $\sigma_\mre$ & Electrolyte ionic conductivity & S/m\\
\hline $\hat{\sigma}_k$ & Effective solid electronic conductivity & S/m \\
\hline $\tdf$ & Thermodynamic factor & -\\
\hline $\alpha_{\mathrm{a}k}$ & Anodic charge transfer coefficient & -\\
\hline $\alpha_{\mathrm{c}k}$ & Cathodic charge transfer coefficient & -\\
\hline $\theta_k$ & Electrode stoichiometry & -\\
\hline $\Theta$ & Full cell state of charge & -\\
\hline
\end{tabular}
\end{table}

\begin{table}
\caption{Subscripts used in the formulation of the DFN model}
\label{subscripts}
\centering
\begin{tabular}{|c|c|}
\hline \bf{Subscript} & \bf{Description}\\
\hline e & in electrolyte\\
\hline n & in negative electrode/particle (anode)\\
\hline s & in separator\\
\hline p & in positive electrode/particle (cathode)\\
\hline $k$ & in domain $k \in \{\mrn,\mrs,\mrp\}$\\
\hline
\end{tabular}
\end{table}

Both the anode and cathode comprise intermingled aggregates of small ($\sim 10^{-5}$ m) active particles which are connected via a sparse network of a binder with conductive solid additives and permeated by a liquid electrolyte. The electrodes have a typical thickness on the order of $10^{-4}$ m and are therefore approximately one order of magnitude larger than the electrode particles. Directly finding solutions to PDEs on such a geometry is prohibitively expensive, but the DFN modelling framework circumvents this problem by posing the equations in a homogenized, multiscale fashion. The small, or microscopic, length scale is associated with individual electrode particles and the large, or macroscopic, length scale is associated with the electrode width in the direction perpendicular to the current collectors. The details and complexity of the geometry are retained in the multiscale equations (albeit in an approximate fashion) in the form of averaged parameters such as porosity and volumetric surface area that describe local geometry of phase microstructure, as well as `effective' transport properties, whose values depend on the materials they describe, as well as the local microstructure of the multiphase medium in which the materials reside. In the remainder of this section, we outline the model and briefly discuss the physical principles that underpin each of the governing equations.

\paragraph{Macroscopic equations} Two main transport processes are captured on the macroscopic scale, namely (i) material balances in pore-filling liquid-electrolyte phases and (ii) electron conduction in the solid networks of active particles, binder and additives that make up the electrodes. Process (i) is captured by:


\be \label{fc1}
\frac{\partial ( \varepsilon_k c_\mre )}{\partial t}+\frac{\partial N_\mre}{\partial x}=0, \qquad 
N_\mre=-\mathcal{B}_k\De \frac{\partial c_\mre}{\partial x}+(1-\tp)\frac{i_\mre}{F}+c_\mre v_0,\\
\label{fc2} \frac{\partial i_\mre}{\partial x}= b_k j_k, \qquad 
i_\mre=-\mathcal{B}_k\sigma_\mre \left[\frac{\partial \phi_\mre}{\partial x}-\frac{2RT \left(1-\tp\right)}{F}\left( \tdf\right) \frac{1}{c_\mre}\frac{\partial c_\mre}{\partial x}\right],
\ee
in which $R$ is the gas constant, $F$ is Faraday's constant, and $T$ is the ambient temperature. The former equation in (\ref{fc1}) is a material balance, which enforces continuity of the anion concentration, $c_\mre$ --- this equals both the salt concentration and cation concentration as a consequence of the local electroneutrality approximation. (It will always be assumed here that the anion of the lithium salt dissolved in the electrolyte is monovalent.) The latter equation in (\ref{fc1}) is a constitutive law for the total molar flux of anions $N_\mre$, which accounts for contributions due to diffusion, driven by concentration gradients, migration, driven by the ionic current density $i_\mre$, and convection, driven by the solvent velocity $v_0$. Typically, convection is neglected, so $v_0$ vanishes. The migration term accounts for excess flux arising from differences in the relative mobilities of cations and anions. By convention this effect is typically parameterized in terms of a cation transference number relative to the solvent velocity, $\tp$, which measures the fraction of electrolyte conductivity contributed by cation mobility; the anion transference number $( 1 - \tp )$ appears in the flux law here. Diffusion is parametrized by the salt diffusivity in free liquid, $D_\mre$.

In both equations (\ref{fc1}), coefficients taking a subscript $k$ transform the material properties of the free liquid electrolyte into effective values attained when the electrolyte permeates a pore structure within domain $k$. Thus, the volume fraction of phase $k$ occupied by electrolyte, $\varepsilon_k$, appears in the material balance, to make $c_\mre$ represent the anion concentration that would be recorded if electrolyte was extracted from the pores within phase $k$ and measured separately. Similarly, the transport efficiency ${\mathcal B}_k$ appears in the flux law, to account for how pore shapes and connectivity lower the apparent salt diffusivity from $D_\mre$, its value in free liquid. Generally, pore networks with lower porosity and greater tortuosity cause ${\mathcal B}_k$ to reduce from its maximum value of 1. Note that no geometric correction is needed for transference numbers; since they inherently describe a ratio of mobilities, corrections for pore geometry cancel out.

The former equation in (\ref{fc2}) is a charge balance in the electrolyte phase, whose form derives from considering the material balances of cations and anions together, under the local electroneutrality approximation, and an assumption that ion fluxes entirely determine the current density through Faraday's law. A source term appears, dependent on the electrode--electrolyte interfacial current density $j_k$, which describes the rates at which electrochemical reactions occur on pore surfaces. In porous electrodes, $j_k$ quantifies the kinetics of intercalation or deintercalation reactions. Since effective reaction rates also depend on the pore surface area contacted by electrolyte, the generation term in the charge balance scales with a second geometric parameter $b_k$, which describes the electrochemically active surface area per unit volume of the pore network in domain $k$. It is worth emphasizing that here, we shall generally adopt the convention of denoting bulk current densities with $i_k$ and interfacial current densities with $j_k$. Observe that by convention, the current $j_k$ is taken to be positive when a reaction is driven anodically, i.e., when positive charges enter the liquid from the solid phase.

The latter equation in (\ref{fc2}) is the MacInnes equation, a modified form of Ohm's law. As well as relating the ionic current density to the gradient of potential $\phi_\mre$ in the electrolyte through its ionic conductivity $\sigma_\mre$, the MacInnes equation also includes a correction to account for concentration overpotential --- the potential drop associated with differences in the free energy of mixing that accompany concentration polarization. This term injects another material parameter, the thermodynamic factor $1+\partial\ln f_{\pm}/\partial\ln c_{\textrm{e}}$, which accounts for possible deviations from ideal Nernstian behaviour.

Note that the DFN model may also be stated in an equivalent form that includes an explicit balance of the Li$^+$ cations. This can replace either the anion balance or the charge balance, as it depends linearly on those two relationships. The form adopted here is computationally most convenient because LiB electrodes are typically anion blocking. Thus the anion balance equation involves no source terms, a simplification which can be advantageous for numerical solutions, either with finite differencing or the finite-element method (FEM). 

Electron conduction in the anode and cathode is captured by:
\be
\label{fc3}
\frac{ \partial ( i_k + i_\mre )}{\partial x}=0,\qquad 
i_k=-{\hat{\sigma}_k}\frac{\partial \phi_k}{\partial x}.
\ee
Here, $i_k$ is the current density in the solid phase within domain $k$, $\phi_k$ is the potential there, and $\hat{\sigma}_k$ is its effective electronic conductivity. As in the electrolyte-phase governing system, the former equation in (\ref{fc3}) ensures charge conservation, whereas the latter is a constitutive relation, stating that electron conduction is Ohmic. The first equation here expresses overall charge continuity across a volume element of domain $k$. Note that this balance equation could be restricted to the solid phase, including a generation term analogous to the one in the charge balance of (\ref{fc2}), but it is easier to handle the total current density $i_k + i_\mre$ analytically because it is solenoidal (divergence-free) as a consequence of electroneutrality. Observe that $\hat{\sigma}_k$ could in principle be deconvolved into contributions from the different electron-conducting phases that make up the electrode and a geometric factor, analogous to ${\mathcal B}_k$, could be included explicitly, but it is not conventional. We believe that this is likely because it is relatively straight-forward to measure $\hat{\sigma}_k$ directly by experiment.

The local interfacial current density $j_k$ is typically determined by reaction kinetics in electrode domains, and taken to vanish in the separator. Thus
\be
j_k (x) &= \begin{cases}
j_\mrn & \text{ if } 0 \leq x \leq L_\mrn,\\
0 & \text{ if } L_\mrn < x \leq L-L_\mrp,\\
j_\mrp & \text{ if } L_\mrp < x \leq L,
\end{cases}\label{fc4}
\ee
with kinetic rate laws for the nontrivial terms determined by considering reaction mechanisms.

 
\paragraph{Macroscopic Boundary and Interface Conditions}
At the extremities of the cell where the electrodes meet the current collectors, there is no flux of anions and no ionic current; the current in the solid network equals that provided or extracted by the external circuit. At the internal interfaces, where the anode and cathode meet the separator, the current density in the solid network must be zero, because the separator is electronically insulating. As such, the macroscopic equations are supplemented by boundary conditions
\be
\label{fc21} i_\mrn|_{x=0} &=& i_\mathrm{app}, \quad N_\mre|_{x=0} = 0, \quad i_\mre|_{x=0} = 0,\\
\label{fc22} i_\mrn|_{x=L_\mrn} &=& 0, \\
\label{fc23} i_\mrp|_{x=L-L_\mrp} &=& 0, \\
\label{fc24} i_\mrp|_{x=L} &=& i_\mathrm{app}, \quad N_\mre|_{x=L} = 0, \quad i_\mre|_{x=L} = 0.
\ee
where $i_\mathrm{app}$ is the current density discharged by the cell (NB: macroscopic currents that leave the positive electrode are positive by convention, so discharge currents are positive; currents that charge the battery correspond to negative values of $i_\mathrm{app}$).

\paragraph{Microscopic equations and boundary conditions}
As can be seen from the Butler--Volmer laws above, interfacial current $j_k$ depends strongly on the concentration of intercalated Li$^+$ at the surfaces of active particles in the solid phase, $c_{k\mathrm{s}}$. Determination of $c_{k\mathrm{s}}$ necessitates the solution of an appropriate transport or reaction model describing the dynamical distribution of Li within the particles, at every location within the macroscopic dimension, $x$. Transport within particles occurs on a smaller, microscopic scale, taken to be a pseudo-second dimension $r$  orthogonal to $x$. We take $r$ to be the radial position within a spherical particle and let $r_k$ be the radius of a particle in domain $k$. 

There is considerable debate about what equations are most appropriate to describe interfacial reactions and solid-state lithium transport within intercalation particles. In the original papers by Newman and coworkers \autocite{doyle1993modeling,Doyle1995,fuller1994simulation}, a spherical form of Fick's second law of diffusion was adopted, but other models, including shrinking-core \autocite{Zhang2007, GLi2020, Srinivasan2004} and Cahn-Hilliard \autocite{Bazant2013}, have been proposed subsequently to handle materials where phase-transformation reactions \autocite{Newman1995, Karthikeyan2008}, rather than diffusion dynamics, govern the distribution of intercalated lithium. Such materials may also require alternative forms of the microscopic kinetic rate laws.

In an elementary intercalation process, lithium ions are assumed to enter or leave the liquid to occupy available sites in the crystal lattice of an intercalation compound, in which intercalated lithium atoms form a solid solution. Interfacial current is governed by a system of kinetic rate laws of the form
\begin{align}
j_k &= j_{k0} \left[ \exp \left( \frac{F \alpha_{\text{a} k}}{R T} \eta_k \right) - \exp \left( -\frac{F \alpha_{\text{c} k}}{R T} \eta_k \right) \right],\label{fc5}\\
j_{k0} &= F K_k \left( \frac{c_{\mre}}{c_{\mre 0}} \right)^{\alpha_{\text{c} k}}
\left(\frac{c_{k\mathrm{s}}}{c_{k}^{\text{max}}}\right)^{\alpha_{\text{a} k}} \left(1 - \frac{c_{k\mathrm{s}}}{c_{k}^{\text{max}}} \right)^{\alpha_{\text{c} k}} ,\label{fc7}\\
\eta_k &= \phi_{k} - \phi_{\mre} - U_k(c_{k\mathrm{s}}).\label{fc6}
\end{align}
Here, $j_{k0}$ is termed the exchange current density, $c_{k\mathrm{s}}$ is a variable representing the concentration of intercalated lithium at the active-particle surface, and the surface overpotential $\eta_k$ expresses the difference between the liquid--solid potential drop and the equilibrium potential $U_k$, the latter being a state function dependent on $c_{k\mathrm{s}}$. The kinetic model involves various parameters as well: $K_k$ is the rate constant for the  intercalation half-reaction, $c_{\mre 0}$ is the reference concentration of anions in the electrolyte at which $K_k$ is measured, $c_k^{\text{max}}$ is the maximum concentration of lithium that can be intercalated into the electrode material, and $\alpha_{\text{a} k}$ and $\alpha_{\text{c} k}$ represent the charge-transfer coefficients for the anodic and cathodic directions of the half-reaction in electrode $k$, respectively. Note that $K_k$ is alternatively known as the reaction rate or exchange coefficient, depending on the model used  \autocite{Chen2020,Costard2020}.

In line with the assumption that intercalated lithium sits in a solid solution, a diffusion model is appropriate. For simplicity we outline a nonlinear spherical diffusion equation for use in the particle dimension, noting that multiple studies have demonstrated its wide efficacy for fitting experimental trends \autocite{krach18}. In this case the microscopic equations and boundary conditions are:
\begin{align}
\frac{\partial c_{k}}{\partial t} &= \frac{1}{r^2} \frac{\partial}{\partial r} \left(r^2 D_{k} \frac{\partial c_{k}}{\partial r} \right), & \quad \text{ in } 0 < r < r_k,\label{fc30}\\
\frac{\partial c_{k}}{\partial r} &= 0, & \quad \text{ at } r = 0,\label{fc31}\\
- D_{k} \frac{\partial c_{k}}{\partial r} &= \frac{j_k}{F}, & \quad \text{ at } r = r_k. \label{fc32}
\end{align}
where $D_k$ represents the Li-ion solid-state diffusivities in electrode $k$, generally dependent on composition and temperature. Equation (\ref{fc30}) is a material balance for intercalated lithium and (\ref{fc31}) is the condition of boundedness of the concentration, ensuring that concentration remains finite at the particle core. Boundary condition (\ref{fc32}) ensures through kinetic model (\ref{fc5})-(\ref{fc7}) that the $\text{Li}^+$ flux leaving the electrolyte balances that entering the particles. We should emphasise the assumption that electrode particles are spherical here. This is often far from being true, but it is unclear that accounting for more complex particle-scale geometries in fact leads to observably different outcomes. More complex geometries have been considered \autocite{ZhangSastry2007,Thornton2003,Franco2013}, but this often entails solving microscopic transport equations in a higher number of spatial dimensions, which is more computationally intensive and undermines the efficiency of the framework.


\paragraph{Initial conditions}
Initial conditions are required for the ionic concentration in the electrolyte, as well as the Li concentration in the negative and positive electrode particles. Assuming an initially equilibrated state at rest, one can set
\bes
c_\mrn|_{t=0} = c_{\mrn0}, \qquad c_\mrp|_{t=0} = c_{\mrp0}, \qquad c_e|_{t=0}=c_{\mre0},
\ees
where $c_{k0}$ is the initial concentration in electrode $k$ and $c_{\mre0}$ is that in the electrolyte. Section \ref{sec:OCV} discusses the challenges in determining appropriate values for $c_{k0}$, often inferred from the cell voltage, and how individual electrode lithiation is linked to the overall cell state of charge.

\paragraph{Full cell potential}
Computed solutions of the full-cell DFN model can be used to compute potentials at the negative and positive current collectors, $V_\mrn$ and $V_\mrp$ respectively, via the relations
\be \label{fc50}
V_\mrn(t)= \phi_\mrn\big\rvert_{x=0}, \qquad V_\mrp(t)= \phi_\mrp\big\rvert_{x=L}.
\ee
Hence the terminal voltage $V$ measured for a full cell is 
\be
V(t)=V_\mrp(t)-V_\mrn(t).
\ee
This is the cell voltage that should be used to make comparisons between the model and experimental data. It is often helpful to bear in mind that the cell potential, $V$, is composed of several parts, namely: (i) the potential drop across the solid parts of each electrode (associated with electron conduction), (ii) the OCVs between each electrode and the electrolyte, (iii) the reaction overpotentials between each electrode and the electrolyte and (iv) the potential drop across the electrolyte (associated with ionic conduction) \autocite{marquis19}. An appealing way to visualise these different contributions is via a transmission-line model \autocite{LiFarrell2019}, in which there are two parallel branches, each containing resistors, representing the conduction in the solid and electrolyte, respectively. These two branches are connected by a number of nonlinear elements in parallel with one another, representing the (de)intercalation reactions that transfer charge between the electrolyte and solid. A sketch of this transmission line interpretation is given in \textcite{castle21}.

\begin{table}
\centering
\caption{Experimental techniques and parameters that they measure}
\label{table3}
\begin{tabular}{|c|c|c|}
\hline \textbf{Technique}  & \textbf{Model parameter(s) extracted}  & \textbf{Sections for discussion} \\
\hline Microscopy & $L_k$, $r_k$, $\varepsilon_k$, $\varepsilon_{\textrm{act},k}$, $\mathcal{B}_k$ & \S\ref{sec: Imaging} \\
\hline EIS & $\mathcal{B}_k$, $K_k$, $D_k$, $D_\textrm{e}$, $\sigma_\mre$, $\tp$ & \S\ref{sec: EIS Geometry}, \S\ref{frequency}, \S\ref{sec:elyteEIS} \\
\hline XRD &  $r_k$  & \S\ref{sec:laserdiff}\\
\hline Hg porosimetry &  $\varepsilon_k$, $\varepsilon_{\textrm{act},k}$, $b_k$  & \S\ref{sec:mercuryporosimetry}\\
\hline BET adsorption &  $b_k$  & \S\ref{sec:BET}\\
\hline Electronic conductivity probe &  $\hat{\sigma}_k$  & \S\ref{sec:eleccondprobe}\\
\hline pseudo-OCV &  $U_k$  & \S\ref{pOCV}\\
\hline GITT &  $U_k$, $D_k$,   & \S\ref{gitt} \\
\hline SC GITT &  $K_k$,   & \S\ref{gitt} \\
\hline CV &  $K_k$, $D_k$   & \S\ref{cvmeth} \\
\hline PITT &  $K_k$, $D_k$   & \S\ref{potentio} \\
\hline Polarisation-relaxation cells &  $\De$, $\tp$  & \S\ref{sec:polarisation-cells} \\
\hline Concentration cells &  $\tdf$, $\tp$  & \S\ref{sec:conccell} \\
\hline Densitometry &  $\overline{V}_\mre$, $\overline{V}_0$  & \S\ref{sec:densitometry} \\

\hline
\end{tabular}
\end{table}

\section{Geometrical parameters}
\label{sec:geomparams}

\begin{figure}
    \centering
    \includegraphics[width=0.45\textwidth]{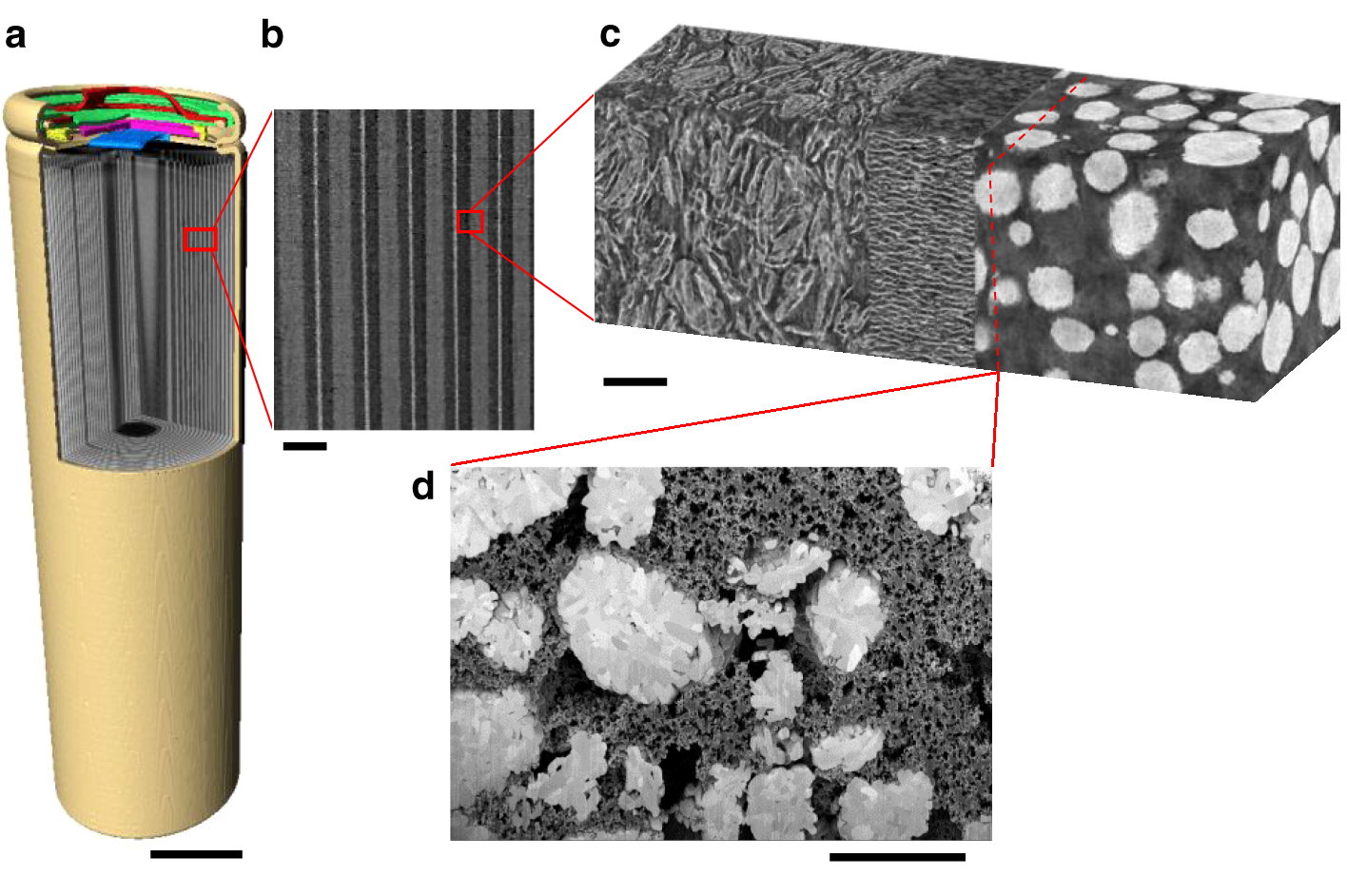}
    \caption{(a) Volume rendering of a reconstructed cylindrical battery scanned by X-ray micro computed tomography. The metal shell (brown), top button (red), venting disk (green), crimp plate (pink), seal insulator (yellow) and current collector (blue) are shown, with a corner cut displaying the internal structure of the battery; (b) magnified virtual slice to show the periodic layered structure of the cell; (c) X-ray nano-CT image showing (from left to right) the graphite anode, polyolefin separator and nickel-manganese-cobalt (NMC) cathode; (d) SEM image showing the carbon-binder domain (CBD) morphology alongside the secondary NMC particles, wherein the crystallography of the primary particles is seen. The scale bars in (a and b) represent 10 mm and 240 $\mu$m respectively, while those in (c and d) are both 10 $\mu$m. Modified from \textcite{Lu2020a}.}
    \label{fig:microstructure}
\end{figure}

The geometric configuration of a cell, as well as the morphologies of the domains within the cell, are determined by manufacturing processes, as well as the inherent structures of their constituents. The choice and balance of components in a given domain determines properties such as active surface area and porosity, and, in turn, these factors significantly influence cell performance. To minimise repetition in the following subsections, it should be stated that although volume fractions, particle sizes, surface areas and tortuosity factors are all typically reported as single, averaged values within each domain (i.e. anode, separator and cathode), they can all vary as functions of position within any domain as well. Renderings of the different layered components are illustrated in figure \ref{fig:microstructure}. Although the various geometrical parameters are discussed in turn, they are intimately linked and in practice it is often not possible to vary them independently. For example, altering average particle radius also alters the surface-to-volume ratio available for electron exchange and effective transport factors.

\subsection{Parameters}

\subsubsection{Layer thickness}

The layer thicknesses of components within the cell, as depicted in Fig.\ \ref{fig:geom}, are straightforward to understand and relatively easy to measure, e.g., using calipers or a micrometre gauge. Electrode layer thicknesses are closely linked to the areal capacities of active material coatings. Reported electrode thicknesses either measured or used in DFN models typically fall in the 10—100 $\mu$m range, and commercial polymer separator thicknesses in the 15—30 $\mu$m range. Layer thicknesses may be affected by the fabrication process, so these properties should be understood as their values after cell formation. This distinction is particularly important for the separator, which is usually made of a porous polymer, whose thickness prior to cell construction can be substantially different from the thickness after being wetted with electrolyte and wound or stacked in a cell, incurring a static mechanical load.

It is also possible that layer thicknesses can change with cycling of a battery, a process that requires further characterization and modelling. These temporal variations are almost always neglected in DFN models, where thicknesses for the electrodes and separator are usually taken to be those measured individually, prior to assembly \autocite{Ecker2015}, or following teardown after use \autocite{Chen2020}. 
Volume expansion or contraction of the active materials accompanies the lithiation/delithiation processes \autocite{Louli2017,foster17}, which is causes concomitant thickening/thinning of the electrodes. For silicon-based anodes, the volume expansion associated with lithiation is so large (up to $\sim$400\% for pure Si) that it cannot be justifiably neglected. In that case some modelling work has aimed to extend the DFN framework to couple volumetric changes with cell electrochemistry, see for example \textcite{McD16} and \textcite{Zhang2020}.

\subsubsection{Volume fractions}

\begin{figure}
\centering
\includegraphics[width=0.7\textwidth]{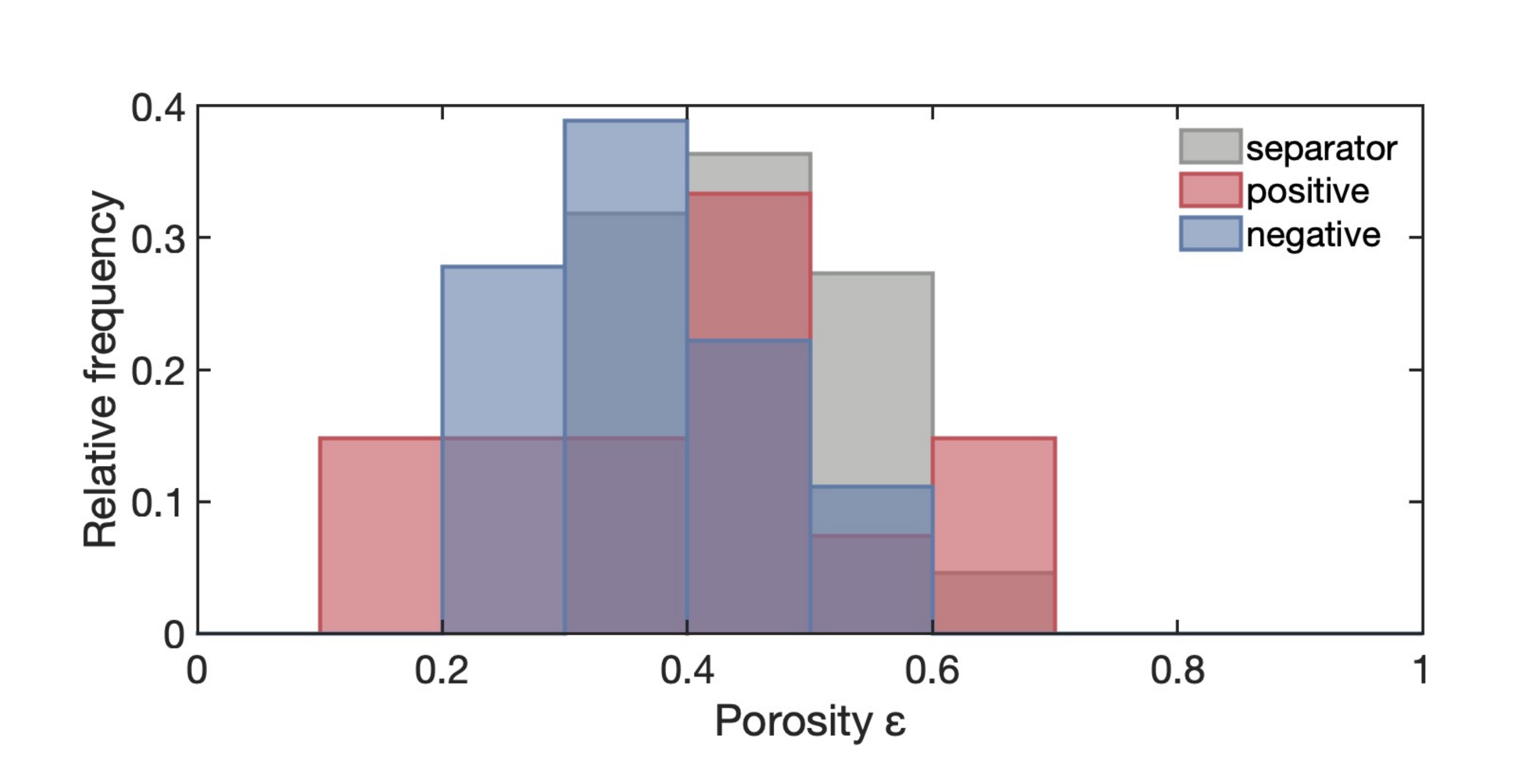}
\caption{Distribution of literature porosities for polymer separators (grey), positive (red) and negative (blue) electrode materials available in the LiionDB parameter database at time of publication \autocite{LiionDB}.}
\label{fig:porosity}
\end{figure}

The two electrodes usually contain four distinct phases: active-material particles, a conductive additive, a polymeric binder, and the pore-filling phase. The conductive additive, most commonly carbon black, is present to increase the electronic conductivity of the composite electrode, aiding electron transfer between the particles and current collector. It is often extremely fine (much finer than the electrode particles) and is typically well mixed with polymeric binder, such that the two phases are hard to distinguish using most imaging techniques. For the purposes of DFN modelling these two phases are usually lumped together and referred to collectively as the carbon--binder domain (CBD). Note that incomplete wetting by electrolyte during cell formation or side reactions during cycling can lead to the formation of a fifth, gaseous pore-filling phase \autocite{Deng2020}, which is typically neglected in DFN models for Li-ion cells, although it has been considered in some cases \autocite{Neidhardt2012}.

Many models, including the original DFN models, set the volume occupied by the CBD to zero \autocite{doyle1993modeling,fuller1994simulation,Chen2020}. This assumes that cells are manufactured to maximise energy density by minimising the required CBD volume. As such, the electrolyte and active material volume fractions are related via:
\begin{equation}\label{eq:volumefrac}
    \varepsilon_k + \varepsilon_{\textrm{act},k} = 1 
\end{equation}
Here, $\varepsilon_k$ is the electrolyte volume fraction (or solid-phase porosity) and $\varepsilon_{\textrm{act},k}$ is the active-material volume fraction. Figure \ref{fig:porosity} illustrates the distribution of porosities for positive and negative electrode materials, along with polymer separators sampled from the LiionDB parameter database. Porous electrodes in LiBs have electrolyte volume fractions of approximately 0.4–0.5, while separators skew towards higher porosities.

Neglecting the CBD volume fraction via the assumption of spherically symmetric electrode particles ignores the fact that parts of an electrode particle's surface will be in contact with the CBD, instead of the electrolyte, resulting in some portion of the particle's surface not being available for (de)intercalation. This sort of consideration could dramatically impact model assumptions, particularly the spherically symmetric diffusion approximation in the microscopic equations and boundary conditions.

The only role that the CBD plays in the model equations presented here is indirect: the CBD volume fraction determines the effective electronic conductivity of the electrodes, $\hat{\sigma}_k$, a quantity that is typically measured directly using electrochemical impedance spectroscopy (EIS). The active-material volume fractions, in combination with the layer thicknesses, are needed to determine the theoretical areal capacity of the cell. They are also used to calculate the maximum Li concentration in active materials, as well as for anode--cathode stoichiometric balancing, as discussed in section \ref{sec:maxconc}. The electrolyte volume fraction has a significant role in determining the rate performance of the cell, because it impacts the transport of ions in the electrolyte required for charge transfer. Further discussion of porosity and its effect on transport efficiency is provided in section \ref{sec:porephasetransport}. On a practical note, incomplete electrolyte infiltration within pores is known to reduce cell performance. Gas generation from degradation during operation may also cause additional inactive regions \autocite{Deng2020, Yang2017}. 

\subsubsection{\label{SAsss}Particle surface area per unit volume}

\begin{figure}
    \centering
    \includegraphics[width=0.7\textwidth]{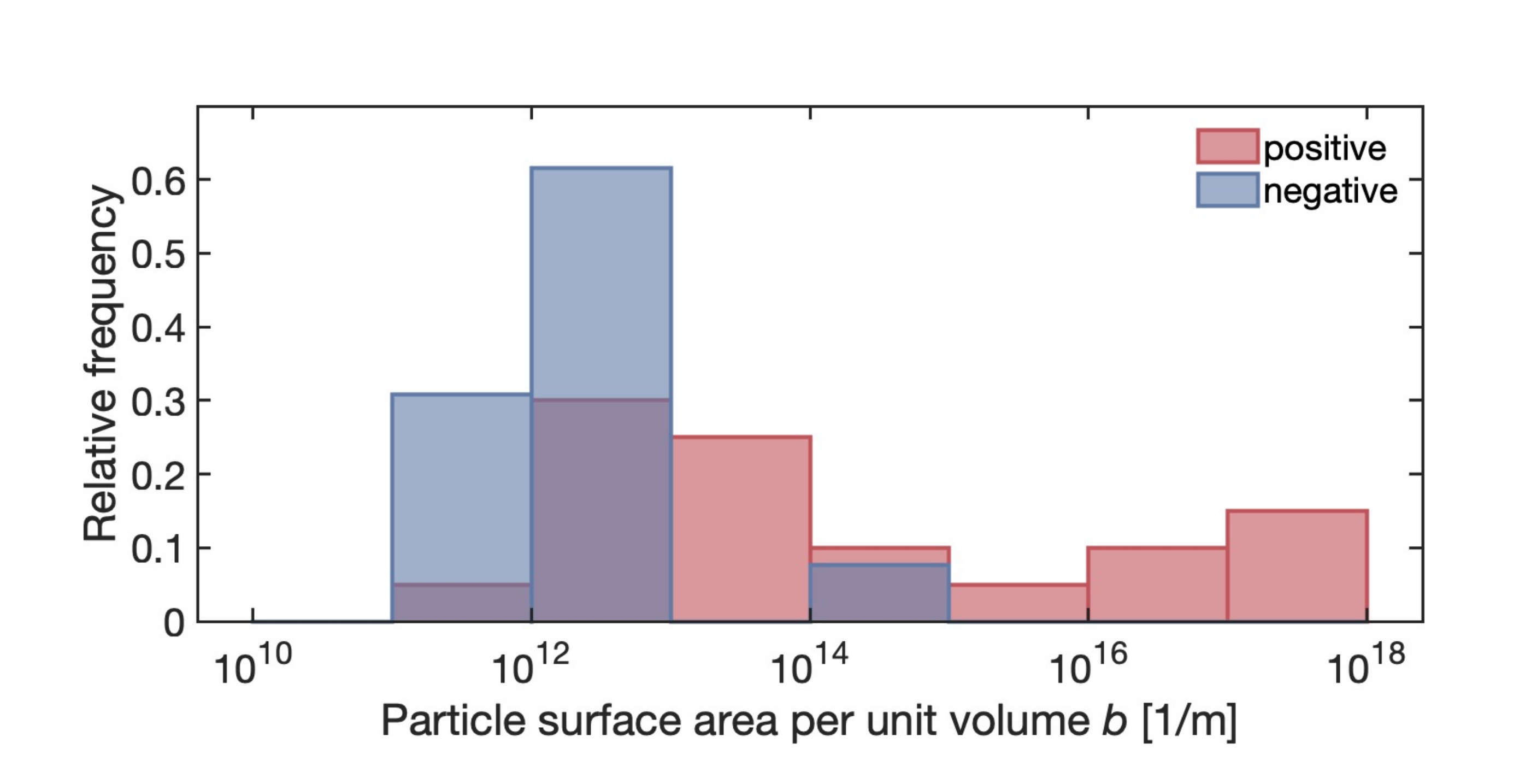}
    \caption{Distribution of literature particle surface areas per unit volume for positive (red) and negative (blue) electrode materials available in the LiionDB parameter database at time of publication \autocite{LiionDB}.}
    \label{fig:surface_b}
\end{figure}

The surface area of active material that is available for intercalation per unit volume of electrode, $b_k$, depends on the size and shape of the active particles as well as their arrangement with respect to the CBD \autocite{Kirk2020}. It comes into the liquid-phase charge balance in (\ref{fc2}) when relating the local interfacial current densities on pore surfaces to the current transferred per unit of porous-electrode volume. As such, the surface-area to volume ratio impacts the rate at which lithium can transfer between the liquid electrolyte and the solid active materials.

Naively, one might expect that $b_k$ should be maximised for the optimal performance (to provide lots of surface area for intercalation); however, in reality, the active material surfaces tend to form a layer called the solid electrolyte interphase (SEI), which consumes lithium inventory, thereby decreasing device capacity, and so a balance must be struck. In addition, the practicalities of manufacturing with very fine powders, which are hard to produce and process, can also be a consideration. 

As highlighted in section \ref{section:geom mft}, and plotted on figure \ref{fig:surface_b}, the measured surface area varies dramatically depending on the measurement technique used. There is also little consensus on the choice of surface area, with literature measurements choosing between definitions such as: (a) the whole solid-pore interface (i.e. including CBD surface), (b) the active particle surface area, and (c) the active particle-pore interface. As stated in section \ref{section:BV Discussion}, the surface area defined must be paired carefully with measured exchange current densities to parameterise electrode kinetics consistently.

\subsubsection{Radius of electrode active particles}

\begin{figure}
    \centering
    \includegraphics[width=0.7\textwidth]{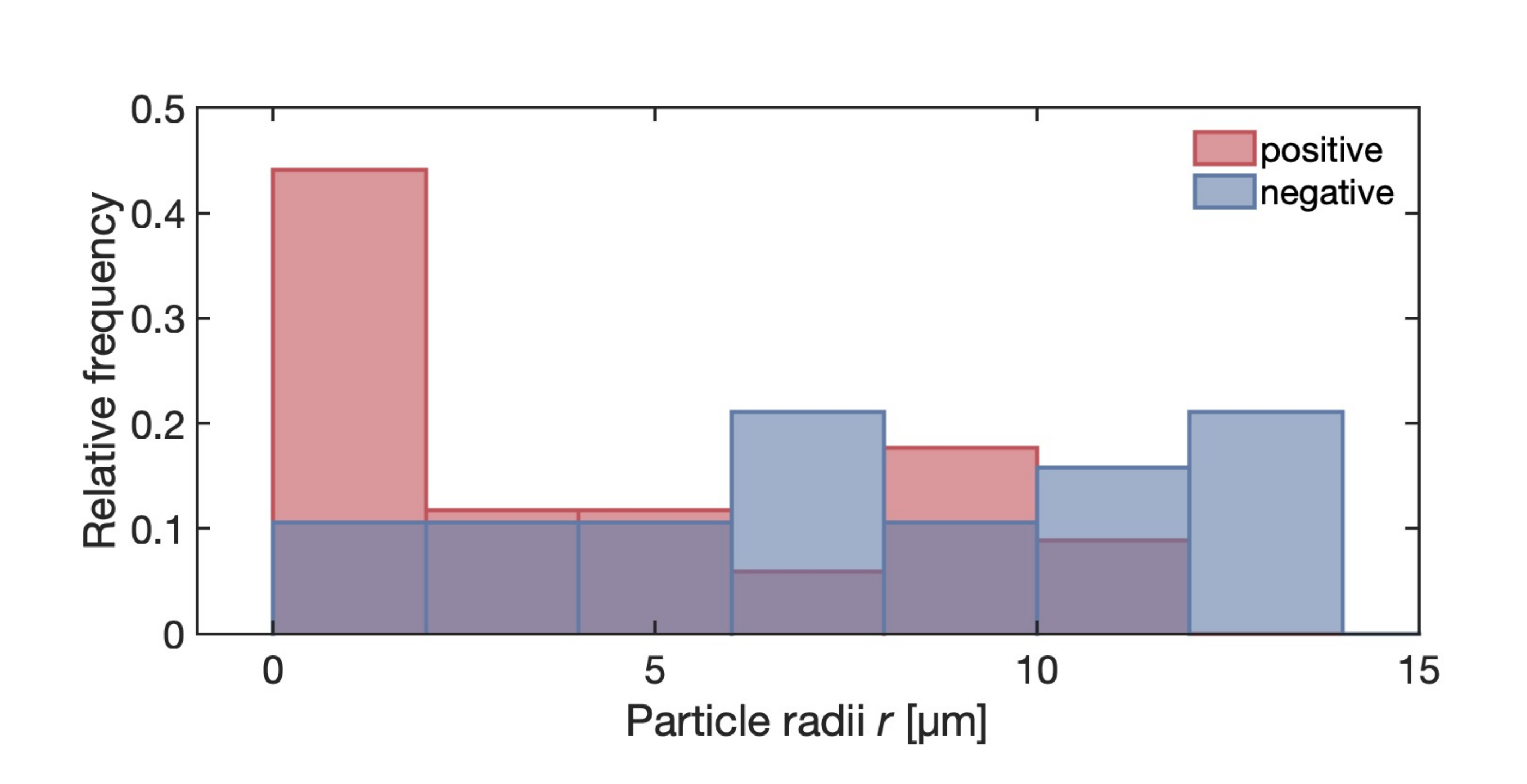}
    \caption{Distribution of literature radii for positive (red) and negative (blue) electrode particles available in the LiionDB parameter database at time of publication \autocite{LiionDB}.}
    \label{fig:particleradii}
\end{figure}

The radii of the active-material particles establish the distance that intercalated lithium need to diffuse to reach the core, as well as determining the characteristic relaxation time of solid-phase diffusion, which in turn controls rate performance. Large particles do not perform well at high rates, because the depletion of or saturation with lithium at particle surfaces limits the accessible range of reaction current, thereby inhibiting performance by lowering power efficiency.

Real electrodes contain a distribution of particle sizes and shapes that confounds the determination of a single representative radius for each electrode, as is required by the basic DFN model \autocite{Cui2016}. For polycrystalline materials, the choice of a characteristic length based on primary or secondary agglomerated structures can change the radius parameter by a factor of 20 \autocite{Ko2019}. Particle radii tend to fall below 20 $\mu$m according to figure \ref{fig:particleradii}. The spread in the literature owes to the choice of primary or secondary structures, along with the variation among active materials. 

The particle radius plays a large role in the way intrinsic parameters of an electrode material are extracted, as discussed in section \ref{sec:electrodeparams}. While average radii are adequate for most models, newly engineered electrodes with graded radii may require models that treat particle radii as a variable with position \autocite{Kirk2020,Hosseinzadeh2018}. Different approaches towards averaging can also alter the effective radii \autocite{Kirk2020}. In addition, some materials, such as graphite platelets, tend to form highly non-spherical particles \autocite{Zhao2018, ZhangSastry2007}. Anisotropic lithium diffusivities also result in different observed behaviours than that expected for an isotropic sphere (particularly in the case of graphite) \autocite{Persson2010}. For these reasons, defining a particle radius to characterise and model electrode geometries is not straightforward, even if all of the true morphological information is known.

\subsubsection{Effective transport properties for electrolytes in pore networks}
\label{sec:porephasetransport}

\begin{figure}
    \centering
    \includegraphics[width=0.7\textwidth]{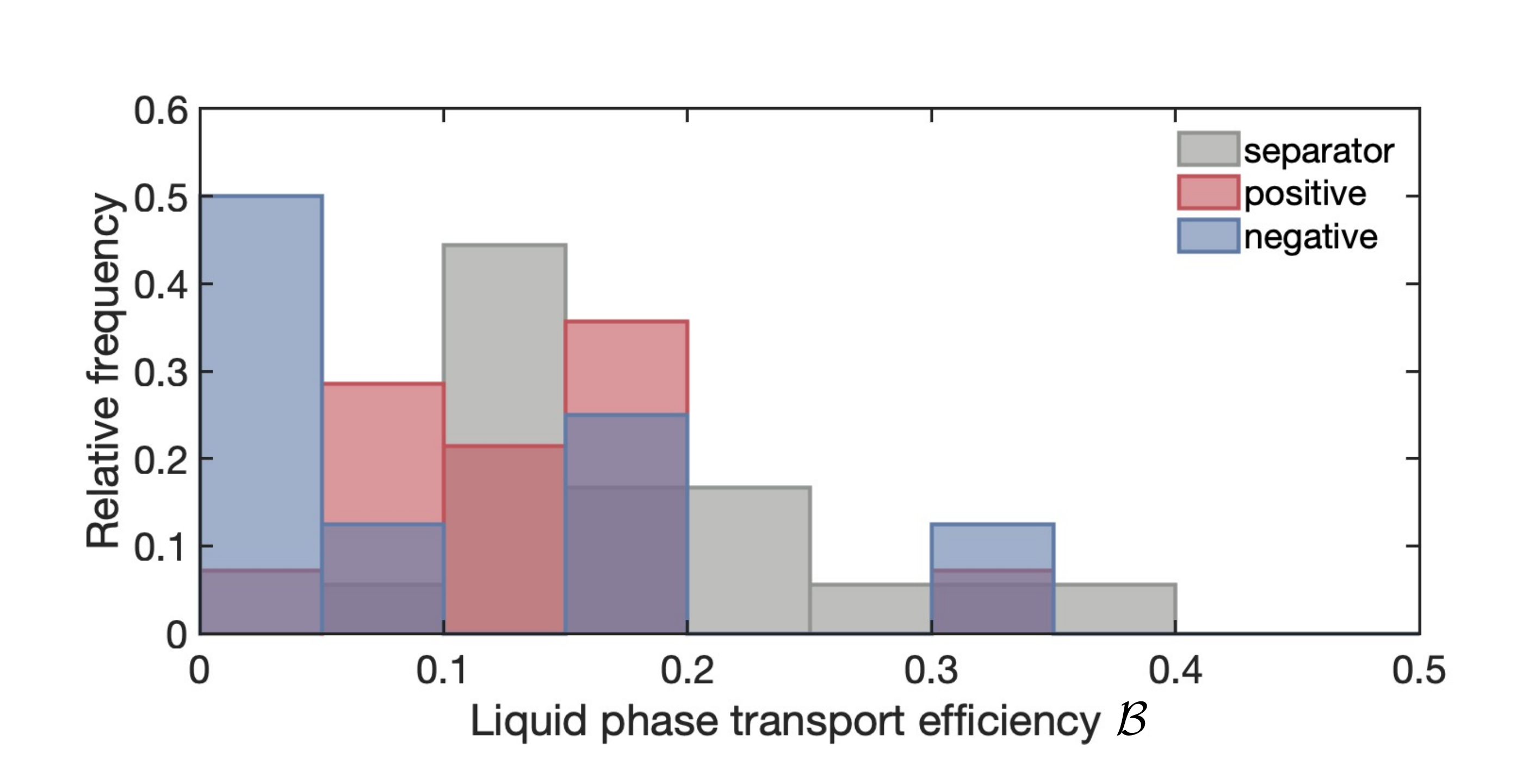}
    \caption{Distribution of liquid-phase transport efficiencies in polymer separators (grey), and positive (red) and negative (blue) electrodes available in the LiionDB parameter database at time of publication \autocite{LiionDB}.}
    \label{fig:curlyB}
\end{figure}

The original DFN formulation captures the impact of pore-network geometry on electrolyte transport properties by using effective transport parameters $D_{\textrm{eff},k}$ and $\sigma_{\textrm{eff},k}$. When a diffusion medium can be suitably homogenized, the domain-specific properties of a pore-filling electrolyte relate to their values in an unobstructed electrolyte 
through a domain-specific transport efficiency, ${\cal B}_k$, as mentioned above. Figure \ref{fig:curlyB} shows that transport through the porous layers in LiBs is typically constricted by more than 60\%. 

%
It is also common to see MacMullin numbers ($N_{\text{M},k}$) being used to quantify the hindrance of transport caused by the electrode geometry \autocite{Landesfeind2016}. As ${\cal B}_k$ --- the inverse of $N_{\text{M},k}$ --- is bounded between 0 and 1, it is comparatively more convenient for manipulation and interpretation. Transport efficiency is often further decomposed into two distinct contributions: the reduction due to the pore volume fraction, captured by $\varepsilon_k$, and the reduction due to the pore tortuosity factor, $\uptau_k$. The four macroscopic factors that characterize pore networks relate through 
\begin{equation}
\label{eq:curlyBtau}
    {\cal B}_k= \frac{1}{N_{\text{M}, k}}=\frac{\varepsilon_k}{\uptau_k}.
\end{equation}
Measurement and interpretation of tortuosity factors has been the subject of extensive discussion \autocite{usseglio2018resolving}. 

\textcite{fuller1994simulation} make use of the well-established Bruggeman correlation (i.e.\ $\uptau_k=\varepsilon_k^{1-b}$, where $b$ is 1.5 for effective transport through spherical packed beds) to approximate the tortuosity factor. We note that there are some subtleties in how \textcite{doyle1993modeling} implement the correction; for example, they take a factor of $\varepsilon_k$ out from $D_{\textrm{eff},k}$, since their variable $N_\mre$ describes anion flux through a whole volume element, whereas their $c_\mre$ is the concentration of anions in the pores.

Bruggeman's tortuosity correlation is commonly deployed in situations where experimental geometrical information is limited. Although it is a convenient means to approximate the effect of microstructure on transport, it has been shown that the assumptions in its derivation do not necessarily align with the conditions established by the microstructures of real battery electrodes \textcite{Tjaden2016}. For example, use of a Bruggeman factor suggests that all electrodes with the same average particle radii and porosity would have the same pore-phase transport properties, but this is clearly not the case. Although correlations can be extended to account for more complex scenarios \autocite{chung2013validity}, many additional simplifications and assumptions are still required.



Unlike porosity, which has a meaningful definition within a planar differential cross section (ratio of cut-through pore area to total plane area), the tortuosity factor (and therefore ${\cal{B}}_k$ and $N_{\text{M},k}$) does not. This is because the tortuosity factor is an emergent property of flow through a microstructure; therefore sufficient path length (i.e. representative volume) is required before a value can be meaningfully measured. The multiscale approach used to construct the DFN model assumes that the characteristic lengthscale of the electrode particles is negligible compared to that of the electrode thickness, but this is not the case, with typical scale differences of only a single order of magnitude. The error that this incurs is likely to be related to the ratio of particle size to electrode thickness, i.e., of the order of 10 \%.

More recently, alternative concepts, including the ``electrode tortuosity factor'' \autocite{nguyen2020electrode} have been reported that are able to measure the effect of graded microstructures. Many literature examples report ``geometrical tortuosities'' or ``path-length tortuosities'', calculated using a wide variety of algorithms applied to image data. This is distinct from the tortuosity factor which is found from using the multiscale approach. However, as made clear by \textcite{epstein1989tortuosity}, although path-length tortuosities are relevant to systems with well-defined paths, e.g., rivers and veins, they are not relevant to the types of microstructure found in real battery electrodes, unless they happen to consist of `capillary' pores with relatively constant cross section.


\subsubsection{\label{effcond}Effective electrode solid conductivity}

In DFN models, the electronic conductivity of the electrodes is often assumed to be constant (with different values within each electrode domain, $\hat{\sigma}_\mrn$ and $\hat{\sigma}_\mrp$). Although the active particles are electronically conductive, they have a much lower conductivity than the CBD. 
Another commonly ignored consideration in continuum-scale models, especially for electrodes that use active materials with low electronic conductivity, is the impact on the effective electrochemically active area of electronically-isolated regions of active material away from the CBD network \autocite{Qi2013}. 

Probe methods capture the effective conductivities directly, and are therefore usually adopted in practice, although two-electrode measurements of direct-current resistance are sometimes used.




\subsection{Measurement/fitting techniques} \label{section:geom mft}


\subsubsection{Imaging} \label{sec: Imaging}

Scanning electron microscopy (SEM) and transmission electron microscopy (TEM) have been used extensively for two-dimensional (2D) image acquisition. SEM scans a beam of electrons across the target in a raster pattern and then detects the scattered electrons, with some devices achieving a resolution of below 1 nm \autocite{goldstein2017scanning}. By contrast, TEM transmits electrons through the sample before being detected, offering a way of capturing the internal detail, such as crystal structure and morphology. Additionally, TEM can achieve finer resolution than SEM, with some reports of 50 pm. However, samples need to be less than $\sim 10^{-7}$ m thick (much thinner than practical electrode coatings) to allow full electron penetration. 

2D images can then be processed to extract measurements for layer thicknesses, particle radii and particle size distributions. It is important to note that because sample sizes are often very small compared to the entire electrode, a given image may not be statistically representative of the whole structure. It is generally difficult to obtain sufficiently large datasets that are truly representative for analysis.

Both focused ion beam-scanning electron microscopy (FIB-SEM) and X-ray computed tomography (XCT) have been used extensively in three-dimensional (3D) studies; many examples can be found in fuel cell \autocite{ostadi20103d,ziegler2011direct}, and battery research \autocite{stephenson2011modeling,le2020effect,finegan2016characterising}. The FIB-SEM method images the 2D surface using an SEM, with the same resolution capability as SEM, and then mills the surface away using a controlled beam of charged positive particles, typically gallium, to reveal another layer. These 2D images are then reconstructed to produce a 3D microstructural volume. The milling process means FIB-SEM is destructive, i.e. \textit{in situ} or \textit{operando} imaging is not possible \autocite{wargo2013comparison}. XCT systems fire an X-ray source at the sample and measure the resulting attenuation pattern, producing a 2D projection image. The sample is rotated to obtain images from many angles, which are then reconstructed, typically through a back-projection algorithm, to produce a 3D volume \autocite{le2021x}. Image analysis methods are used to directly characterise the 3D microstructure, helping to understand particle distribution and anisotropy \autocite{Ebner2013_anisotropy,kehrwald2011local,ebner2013visualization}. 

Image-based simulations can calculate geometric parameters, such as effective conductivities and volume fraction. Several open-source tools and methods for data processing are available \autocite{le2020physics}. Two popular options are (i) TauFactor, an open-source MatLab application for efficiently calculating the tortuosity factor, volume fraction, surface area and triple phase boundary density from image-based microstructural data \autocite{cooper2016taufactor} and (ii) OpenPNM, an open-source software package developed in Python for performing physics-based pore network simulations \autocite{gostick2016openpnm}. Other image-based mesostructure simulations also account for the CBD in electrode models, to elucidate effective transport properties and extensions into mechanical stresses \autocite{Trembacki2017,Chouchane2019}.


\subsubsection{Electrochemical impedance spectroscopy}  \label{sec: EIS Geometry}

Electrochemical impedance spectroscopy (EIS) is an electrical technique useful to probe various phenomena within LiBs \autocite{Orazem2008}. This section describes the technique but focuses on geometric parameters; sections \ref{frequency} and \ref{sec:elyteEIS} will address material and interfacial properties.

The EIS method consists of applying a small sinusoidal voltage or current perturbation and recording the respective current or voltage response, whose resulting amplitude and phase shift quantify the impedance; an impedance spectrum is then recorded as a function of the frequency of the excitation signal. Although frequencies ranging from $\mu$Hz to MHz are in principle accessible, most information is gathered in the range between 0.1 Hz and 100 kHz, where signal-to-noise ratios are good and reproducibility is easier to achieve. EIS provides multiscale resolution, because different physicochemical processes have distinct characteristic timescales, excited at different frequencies within the sampled spectrum. 

Diffusive transport --- both in electrode active particles and the electrolyte --- has characteristic frequencies in the range below 1 Hz. Interfacial effects, such as charge-transfer resistance (kinetic rate constants), double-layer capacitance, and possibly effects arising from surface passivation are resolved at intermediate frequencies, between 1 Hz and 1 kHz. At higher frequencies, above 1 kHz), ohmic and ionic resistance and inductance can be observed. Notably, the linkage between impedance data and material or geometric parameters requires an appropriate model be fitted to the observed spectrum. Often the choice is an ECM whose parameters relate heuristically to physical parameters \autocite{Li2014,Wang2018}, but formally linearised electrochemical models may also be used \autocite{Doyle2000, Bizeray2019}.

Generally, division of the effective electrolyte conductivity extracted from EIS by the ionic conductivity independently measured for unobstructed liquid directly yields the transport efficiency $\mathcal{B}_k$ discussed in section \ref{sec:porephasetransport}.
\textcite{Landesfeind2016} surveyed many specific details about methods by which Macmullin numbers or transport efficiencies can be determined. Using fits of EIS data with transmission-line models, they determined transport efficiencies in a range of porous electrodes and compared these with values calculated from the Bruggemann correlation typically used in DFN models. 

\subsubsection{Laser diffraction particle size analysis} \label{sec:laserdiff}

Laser diffraction particle sizers that can measure the distribution of particle radii within powders are commercially available. Typically they use photodetectors to measure the angle and intensity of light scattered through a sample. The collected pattern is then transformed into a particle-size distribution via the Mie theory of scattering. It is common to describe the spread in size distribution with D$_{10}$, D$_{50}$ and D$_{90}$ diameters, representing the size at the 10th, 50th and 90th percentiles, respectively. This technique has been widely applied to active-material samples \autocite{ebner2013x,Drezen2007,Cui2016}. A flaw of particle-size measurement via laser diffraction is that it assumes scattering from perfect spheres and generally does not provide accurate distributions from the irregularly shaped particles that make up some battery materials \autocite{Eshel2004}.

\subsubsection{Mercury intrusion porosimetry} \label{sec:mercuryporosimetry}

Mercury intrusion porosimetry is a destructive technique used to measure porosity, volume of pores, pore size distribution and surface area. In a measurement, increasing pressure is applied to force mercury into smaller and smaller void spaces in the porous medium being studied, and recording the volume of intruded mercury \autocite{Giesche2006}.

Smaller pores may not always be fully accessible to mercury intrusion porosimetry, owing to the higher pressures necessary \autocite{sheng2014effect}. As such, the method is only suitable for battery electrode coatings with larger pore sizes \autocite{Ecker2015,Chen2020,Schmalstieg2018}. It can also be used to measure particle radii within electrodes by exploring the contact angle, hydrostatic pressure, and surface tension of the mercury \autocite{Schmalstieg2018}. Similar to laser diffraction, the principles of mercury intrusion porosimetry derive from the assumption of perfectly packed spheres \autocite{Mayer1965}. Geometrical parameters for inhomogenous active material particles may be more precisely extracted using imaging methods.

\subsubsection{Brunauer–Emmett–Teller adsorption} \label{sec:BET}
The Brunauer–Emmett–Teller (BET) adsorption isotherm can be used to measure surface areas within porous media by flushing the pore structrure with nonadsorptive gas such as helium, then inducing the adsorption of a surface monolayer of an adsorptive gas, such as nitrogen \autocite{Sinha2019}. Based on formulas derived from the BET theory, the quantity of gas adsorbed correlates with the total surface area of the particles. By tracking the hysteresis of nitrogen adsoprtion and desoprtion, experimental equipment that performs measurements with the BET method can often also be used to infer pore-size distributions.

BET adsorption is a prevalent tool, widely used to extract solid electrode-material surface areas under varying synthesis conditions and degradation/cycling protocols \autocite{Cho2001,huang2014micro,Lim2008,Winter1998}. Shortcomings of the technique include the high sensitivity to sample preparation, degassing and lengthy measurement time. BET adsorption cannot deliver very accurate parameters for DFN models, because the technique can only be used on dry powders, and therefore does not characterise the actual areas in chemical contact or available for electron exchange within porous electrodes.


\section{Electrode parameters}
\label{sec:electrodeparams}
The anode and cathode active materials undergo redox reactions, contribute towards electron conduction and, most importantly, serve as reservoirs for lithium storage. Typical cathodes comprise transition-metal oxides, such as blends of lithium nickel cobalt manganese oxides, or phosphates, such as lithium iron phosphate (LFP). Anodes typically contain carbonaceous compounds such as graphite, with some incorporating varying amounts of silicon. This section focuses on electrode intrinsic parameters encompassing thermodynamic, kinetic and transport properties, that are independent of pore geometry. 

\subsection{Parameters}

\subsubsection{Open-circuit voltage (OCV)} \label{sec:OCV}

\begin{figure}
    \centering
    \includegraphics[width=0.8\textwidth]{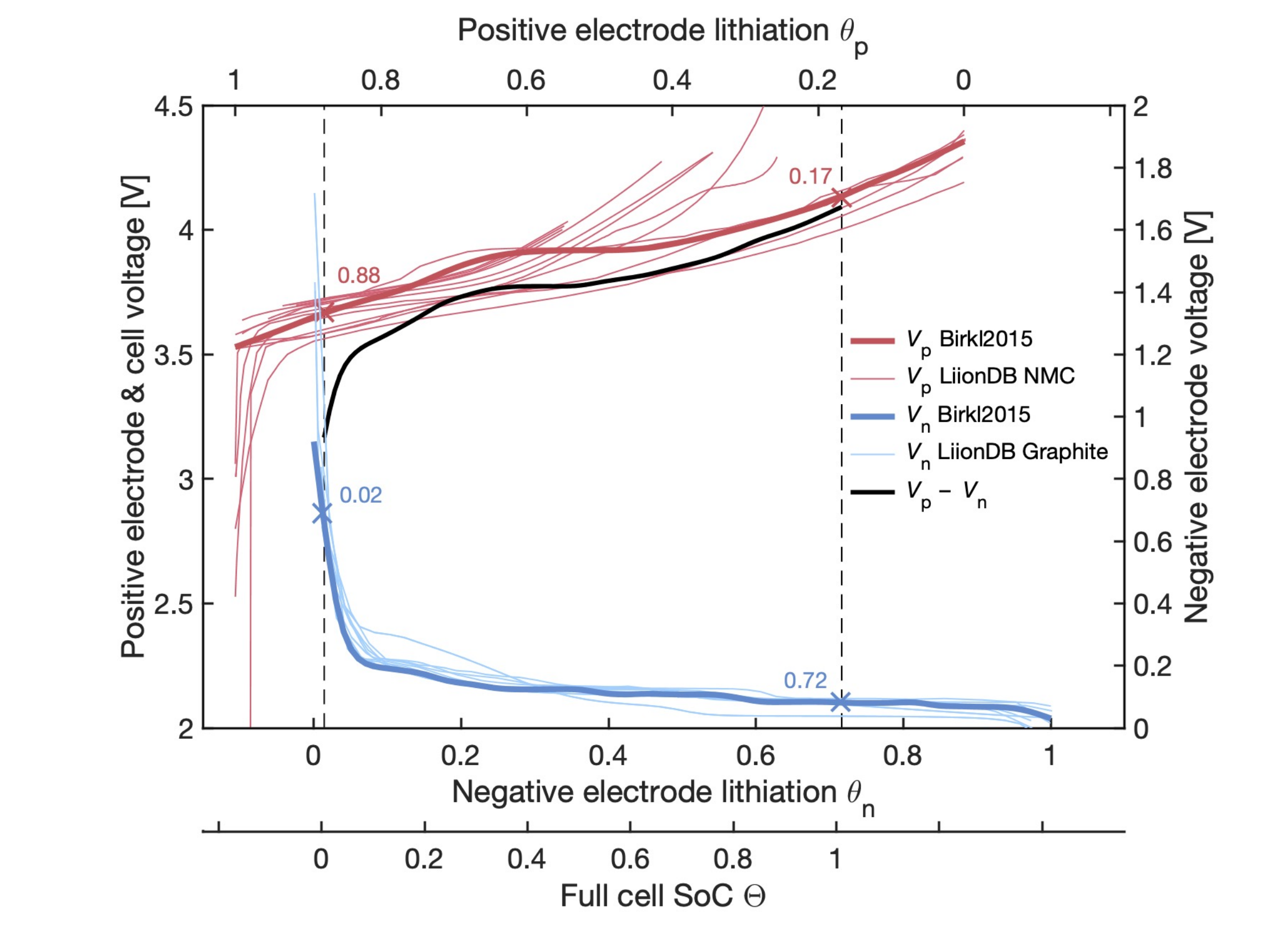}  
    \caption{Equilibrium open-circuit voltages(OCVs) for  nickel-manganese-cobalt (NMC) cathodes (blue —), graphitic anodes (red —) and combined full cell (black —). Thin lines represent NMC and graphitic material OCVs, as reported by other literature sources collected in the accompanying database \autocite{LiionDB}. Stoichiometric limits of the cathode and anode are indicated by the vertical dashed lines (---) and markers (x) on electrode-specific axes.}
    \label{fig:OCV_Balancing}
\end{figure}

The OCV of a full cell is defined as the potential difference between the positive and negative electrodes when there is no current flow and the cell has relaxed to equilibrium. The OCV is not a constant, but is instead a function of the thermodynamic states of the materials in the electrodes. Generally, the OCV varies significantly with respect to its degree of lithiation (i.e.\ its SoC), and also slightly with temperature.

To characterise the full-cell OCV, the variation of potential of each of the active materials at equilibrium is measured, both relative to the same reference-electrode reaction \autocite{Guggenheim1929} --- most commonly lithium redox on lithium metal --- taking the difference between these two voltages (often termed half-cell OCVs) gives the full-cell OCV. This data can then be supplied to the model either in the form of a look-up table or a suitable mathematical function fitted to the raw data. OCV data is commonly fitted to either (i) functions that are empirical but convenient to work with (e.g., piecewise linear functions \autocite{Reniers2019}), see e.g. \autocite{Dees2008}, or (ii) physically motivated functions \autocite{birkl2015parametric}. Some of the challenges associated with data-fitting in practice have been discussed by Plett and colleagues \autocite{plett1,plett2}. The original modelling work by Newman and co-workers used either a form of the Nernst equation \autocite{doyle1993modeling} or an empirical expression including hyperbolic tangent terms, exponentials and a constant \autocite{fuller1994simulation}. Similar approaches have been widely adopted in many other more recent works \autocite{Farkhondeh2013,Verma2017,Srinivasan2004}. Some authors have also fitted a polynomial to the measured data, which may be sufficient for many applications and is simpler to implement \autocite{zhao2017observability}.

\textcite{birkl2015parametric} and \textcite{verbrugge2017thermodynamic} both demonstrated that a realistic function for each half-cell OCV can be derived from first principles using a substitutional lattice gas model. The derivation of these functions is similar to that of a Nernst equation, but is extended to include the energetics of short-range interactions and local equilibria among sites with different energies. The model was demonstrated to accurately represent a wide range of common materials, including graphite, layered nickel-manganese-cobalt oxide, manganese oxide, and iron phosphate. The function is a summation of sigmoidal terms for each plateau in the OCV. In Birkl's work \autocite{birkl2015parametric}, the fraction of sites within the active material occupied by lithium, $\theta_k$, is given in terms of the OCV, $U_k$, by 
\begin{equation}
    \theta_k(U_k) = \sum_{i=1}^N \frac{\Delta\theta_i}{1 + \exp{\left[ \left(U_k - V_{0,i}\right) a_i e / kT\right]}}.
    \label{birklocv}
\end{equation}
The voltage of each plateau is $V_{0,i}$; the fraction of the SoC range in which the material resides on the $i$th plateau is $\Delta \theta_i$, and $a_i$ is a fitting parameter that approximates the interaction energy between neighbouring ions in the lattice. Verbrugge's expression is derived in a similar manner from equilibrium thermodynamics \autocite{verbrugge2017thermodynamic}, but includes a lattice-energy parameter equivalent to the reciprocal of Birkl's $a_i$. Being analytic, (\ref{birklocv}) is smooth and is therefore apt for use in numerical computations. Unlike the familiar Nernst equation, however, the composition is given explicitly, rather than voltage being an explicit function of the composition. Although expression (\ref{birklocv}) is generally invertible because $\theta_k$ always decreases monotonically with $U_k$, the inverse function $U_k(\theta_k)$ cannot generally be found analytically, posing a minor implementational inconvenience.


Hysteresis or path dependence effects are usually observed in OCVs, manifesting as a different OCV function being measured during charge, compared to discharge. The reasons for this are widely debated, but generally accepted to be complex and linked to the formation of metastable states within active materials. Hysteresis may be examined using isothermal calorimetry \autocite{assat2019probing}, and has been modelled empirically by constructing a differential equation that describes how the hysteresis voltage changes with respect to SoC and history \autocite{plett2015battery}.



\subsubsection{Stoichiometric limits and electrode balancing}
\label{sec:maxconc}
The maximum lithium concentration, $c^{\max}_{k}$, dictates the extent to which a solid electrode material can incorporate lithium. Materials that can store large amounts of Li per unit weight or volume are generally sought after, since they increase the energy density of the battery. The parameter $c^{\max}_{k}$ is material-dependent and typically lies between 20,000--55,000 mol/m$^3$ for metal-oxide cathodes, and 15,000--35,000 mol/m$^3$ for graphitic anodes. This parameter appears in model equation (\ref{fc7}), defining the Butler–Volmer reaction rates. Notably, for intercalation compounds that form solid solutions, the exchange current density, $j_{k0}$, vanishes either when the lithium concentration at the electrode surface approaches its maximum or when it reaches zero. 

The theoretical capacity and $c_k^{\text{max}}$ for material $k$ may be calculated, through Faraday's law, by using the active material crystal density, $\rho_k^\text{crystal}$ and the molecular mass of the lithiated active material, $m_k$. \textcite{Chen2020} estimated similar values of $c_k^{\text{max}}$ using both the material coating mass per unit area $M_{\text{coat},k}$ and active material crystal density, through 
\begin{equation}
    c_k^{\text{max}} = \frac{\rho^\text{crystal}}{m_k}
    \approx
    \frac{M_{\text{coat},k}}{m_kL_k\varepsilon_{\textrm{act},k}}.
    \label{eq:maxconc_chen}
\end{equation}


Individual electrodes can typically access maximum and minimum lithium concentrations that lie beyond the capacity ranges accessed in a full cell, as depicted in Figure \ref{fig:OCV_Balancing}. Most positive electrode materials, such as LiCoO$_2$ and nickel-manganese-cobalt oxide, LiNi$_x$Mn$_y$Co$_z$O$_2$ (NMC), undergo irreversible changes of their crystal structure at low stoichiometry, which should be avoided to minimise degradation. Full lithiation of the positive electrode is also not usually possible because cells are fabricated with fully lithiated cathode materials; some capacity is inevitably lost due to the loss of lithium inventory during initial SEI formation. For the negative electrode, it is important to leave some margin to avoid overcharging or overdischarging a cell. 

The literature is often unclear whether stoichiometry is defined in terms of (i) a fractional of the cell's SoC (ranging between 0 and 1) that is accessible in practice, but not necessarily between 0 and $c_k^{max}$, or (ii) the true theoretical maximum. Adopting a perspective in which lithiation limits are imposed, the fractional SoC $\Theta(t)$ can be defined as
\begin{equation}
    \Theta(t) = \Theta(0)-\frac{A\int_0^t i_\mathrm{app}(t^\prime)\,\mathrm{d}t^\prime}{Q_\mathrm{cell}},
\end{equation}
where $Q_\mathrm{cell}$ is the capacity of the cell when fully charged (but the anode material may not necessarily be fully lithiated and, likewise, the cathode material may necessarily not be fully delithiated), so that $\Theta>1$ corresponds to overcharge and $\Theta<0$ corresponds to overdischarge. Each electrode's state of lithiation is expressed with a variable ($\theta_k$ in equation \eqref{birklocv}) which determines the fraction of crystal-lattice sites occupied by Li$^+$. Since $\theta_k$ varies in space as well as over time, it is useful to define a volume-averaged fraction $\theta_k(t)$ of the maximum concentration as
\begin{equation}
    \theta_k(t)=\frac{\int c_k(x,r,t)\,\mathrm{d}V}{Vc_{k,\text{max}}}=\frac{3\int\int_0^{R_k} c_k(x,r,t)r^2\,\mathrm{d}r\,\mathrm{d}x}{L_kr_k^3c_{k,\text{max}}}.
\end{equation}
Thus the full-cell SoC $\Theta(t)$ and the half-cell lithiation fractions, $\theta_\mrn(t)$ and $\theta_\mrp(t)$, are linked by
\begin{equation}
    \Theta(t) = \Theta(0) + \frac{FA}{Q_\text{cell}} \left\{ L_\mrp c_{\mrp,\text{max}}[\theta_\mrp(t)-\theta_\mrp(0)] - L_\mrn c_{\mrn,\text{max}}[\theta_\mrn(t)-\theta_\mrn(0)] \right\}.
\end{equation}
Observe that the electrode-domain thicknesses $L_\mrp$ and $L_\text{n}$ appear here. One important use of DFN models is to optimize electrode balancing by illustrating how choices of $L_\mrp$ and $L_mrn$ affect performance.

Because the full range of individual electrode stoichiometries, $0<\theta_\mrp <1$ and $0<\theta_\mrn <1$, extends beyond the accessible full-cell SoC range, $0<\Theta<1$, full-cell OCV measurements are not sufficient to determine $\theta_k$; in principle half-cell thermodynamic data from the literature or half-cell measurements with harvested electrodes are also necessary. To match half-cell OCV curves with full-cell OCV data, one can examine turning points in differential voltage measurements \autocite{Ecker2015}, perform least-squares regression \autocite{Schmalstieg2018}, or implement three-electrode-cell measurements \autocite{birkl2017degradation,Chen2020}. The initial values of $\theta_k$ relate to the mass balance between the anode and cathode, the first cycle loss observed during the formation process (usually taking place during manufacturing) and the state of health.

Figure \ref{fig:OCV_Balancing} illustrates the range of OCV curves reported in the literature for graphitic anodes and NMC cathodes. There is a lack of consensus between reporting of cell-level SoC, $\Theta(t)$, and materials-level lithiation scales, $\theta_k(t)$, shown particularly by the variation in lithium stoichiometry and measured half-cell OCV for NMC cathodes. Figure \ref{fig:OCV_Balancing} also highlights an example of electrode stoichiometric balancing, from the work of \textcite{birkl2015parametric}, showing how the relative positions of half-cell OCV curves and their minimum/maximum lithiation limits combine to form the full-cell voltage curve. The state of charge at cell level, $\Theta(t)$, and the maximum lithium concentration at the materials level, $c_{k,\text{max}}$ have clear, consistent definitions across the literature, but the variable $\theta_k(t)$ that links the two does not, and is often defined differently in different papers. 

\subsubsection{Butler–Volmer reaction rate constant} \label{section:BV Discussion}

\begin{figure}
    \centering
    \includegraphics[width=0.7\textwidth]{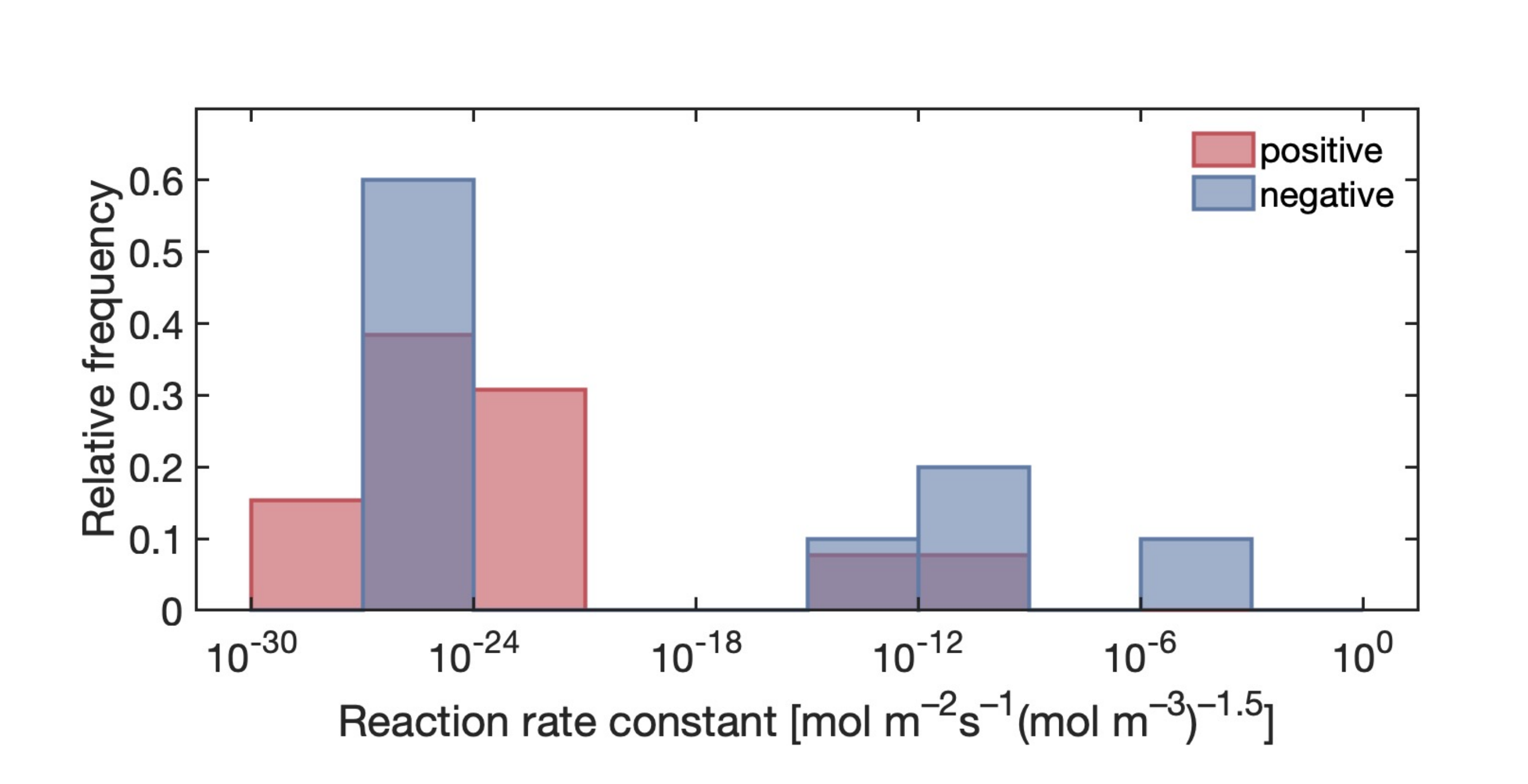}
    \caption{Distribution of literature kinetic rate constants and positive (red) and negative (red) electrode materials available in the LiionDB parameter database at time of publication \autocite{LiionDB}.}
    \label{fig:rateconstant}
\end{figure}

Reaction kinetics in porous electrodes are typically taken to follow the Butler--Volmer kinetic equations (\ref{fc5})-(\ref{fc7}), which relate the interfacial current density $j_k$ to the surface overpotential $\eta_k$. The exchange current density $j_{k0}$, and hence the reaction rate, is proportional to the Butler--Volmer kinetic rate constant, $K_k$. A sample distribution of kinetic rate constants for electrode materials reported from the LiionDB parameter database are shown on figure \ref{fig:rateconstant}. Experimentally obtained exchange current densities vary by several orders of magnitude, with some variation owing to how they are defined \autocite{Bai2014,Smith2017}. As highlighted by Dickinson and Wain, this variability may stem from differences in how variables in the Butler--Volmer equation are interpreted by electroanalytical chemists and electrochemical engineers \autocite{Dickinson2020}. Differences also arise due to parameter values assumed when processing experimental data, such as effective electrode surface area, discussed in section \ref{sec:geomparams}. Caution is advised when comparing published values of $j_{k0}$ and $K_k$, since the Faraday constant, $F$, may be omitted from the definition of the interfacial or exchange current densities in some papers, and the effective surface area may not be taken into consideration \autocite{doyle1993modeling,Albertus2009,Srinivasan2004}.

\textcite{Ecker2015} found that equation (\ref{fc5}) was a good fit to exchange current densities measured by EIS for LiNi$_{0.4}$Co$_{0.6}$O$_2$ (NMC406), but not for graphite. \textcite{Schmalstieg2018} also found that (\ref{fc5}) was a good fit to exchange current densities measured by EIS measurements for NMC811, but the exchange current density had virtually no SoC dependence for graphite. These findings are consistent with SEI-layer properties interfering with the measurements on negative electrodes. Conversely, the same paper by \textcite{Ecker2015} found that the Butler--Volmer equation (\ref{fc4}) provided a good fit for EIS measurements of charge-transfer resistance on graphite, but not for NMC406. Similarly, \textcite{Ko2019} found that $K_p$ extracted from galvanostatic intermittent titration (GITT) measurements on NMC111 is not constant, but instead varies with SoC. Recent literature has also suggested that alternative kinetic mechanisms better describe electrochemical reactions at high rates and overpotentials, or with advanced battery formats, such as lithium metal and anode-free cells \autocite{Bai2014,Smith2017,Sripad2020,Kurchin2020}. A rigorous justification for the common assumption that $\alpha_{\text{a} k} = \alpha_{\text{c} k} = 0.5$ is also lacking \autocite{Newman1975,OKane2020}, but the factors $\alpha_{\text{a} k}$ and $\alpha_{\text{c} k}$ are notoriously difficult to measure. Deviations in these factors from 0.5 has been used to explain 
the fact that LiB discharging is faster than charging at a given magnitude of cell voltage \autocite{Li2019}. 

Thermal activation of intercalation kinetics in materials such as graphite and NMC is widely accepted, with Arrhenius relationships assumed between exchange current density and temperature having been reported throughout the literature \autocite{Ecker2015,Schmalstieg2018,Yang2019,Chen2020}. Despite being called into question from both theoretical and experimental perspectives, the theory of thermally activated Butler--Volmer kinetics is almost always used in LiB models, for two reasons. First, it requires only two parameters: the rate constant in a reference state, $K_k$, and associated activation energy. Second, sensitivity analyses \autocite{Park2018,Dufour2019,Li2020a} have shown that DFN models are more sensitive to particle radii, $r_k$, and diffusivities, $D_k$, than to reaction kinetics.

\subsubsection{Solid-state diffusivity}

\begin{figure}
    \centering
    \includegraphics[width=0.8\textwidth]{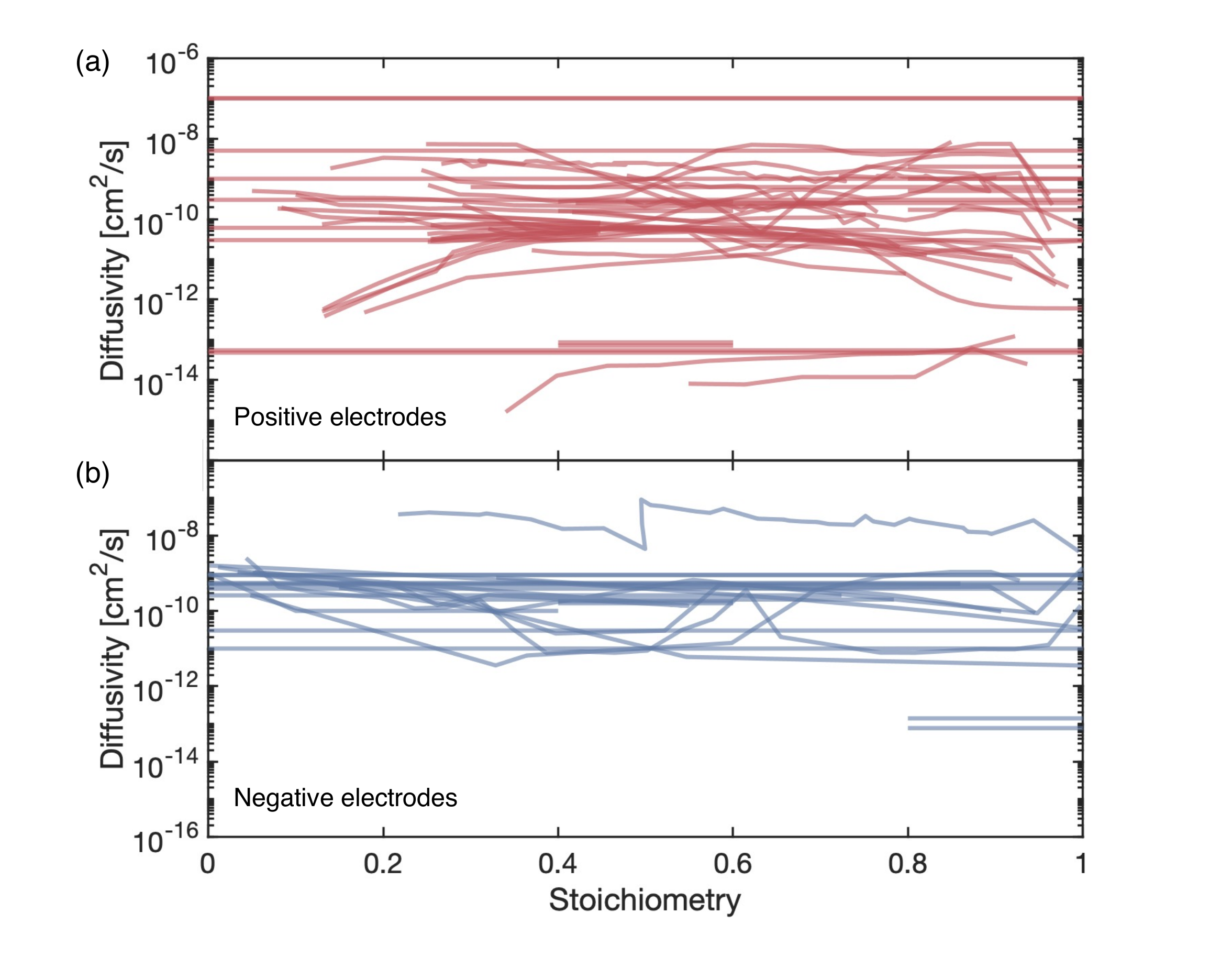}
    \caption{Lithium diffusivity for cobalt containing positive electrodes, $D_\textrm{p}$ (a), and graphitic negative electrodes, $D_\textrm{n}$ (b), as a function of lithiation at 25 $^{\circ}$C. Diffusivities were extracted from literature and reported in the LiionDB parameter database at time of publication \autocite{LiionDB}.}
    \label{fig:soliddiff}
\end{figure}

In the active materials, the Li$^+$ diffusivity, $D_k$, controls the transport of Li$^+$ within the particles. It becomes very important at high currents, where concentration gradients within the active material may become the main source of overpotential if diffusivities are not large enough. When diffusivities are too small, particle surfaces easily become depleted of or saturated with lithium, which shuts down reaction kinetics and thereby strongly impacts battery operation \autocite{Park2018,Dufour2019,Li2020a}.

The first DFN models assumed that solid phase diffusion follows Fick's second law \autocite{Doyle1995}, i.e. they used (\ref{fc30}) and took $D_{\text{k}}$ to be constant. 
It should be emphasised that many of the experimental techniques, such as GITT, infer diffusivities under the assumption that a spherical form of Fick's second law applies and so this model assumption would be required for consistency.

Despite the constancy of $D_k$ assumed in the original DFN models, measurements indicate that $D_\text{k}(c_k)$ can vary by orders of magnitude, as the Li$^+$ concentration changes \autocite{Levi1997, VanderVen2012,Baker2012,Wu2012,Ecker2015,Schmalstieg2018,Chen2020}. The data plotted for positive and negative electrodes in figure \ref{fig:soliddiff} show that Li$^+$ typically diffuses through the active materials in the range of $10^{-8}$ to $10^{-14}$ cm$^2$s$^{-1}$. For graphite negative electrodes, $D_\mrn(c_\mrn)$ is large for concentrations corresponding to material phase transitions between plateaus on the half cell OCV curve and is small for other stoichiometries. Active materials are typically polycrystalline and both modelling \autocite{Han2013} and experimental \autocite{Yang2016} studies on NMC-type materials agree that diffusion is faster along grain boundaries than within crystals. Cracks form much more quickly during cycling in polycrystalline materials than in single crystal materials, however, causing a further decrease in diffusivity \autocite{Yang2016,Trevisanello2021}. Nanoscale LFP shows very fast Li-ion transport, but larger LFP crystals have much slower transport, owing to the transport along the rapid b channels becoming blocked by dislocations. Thus, active material quality has a significant impact on battery behaviour.

Concentration-dependent diffusivities have been used in models, which have been shown to match experimental data better at various C-rates and temperatures \autocite{Ecker2015}. They have also been used to investigate how phase changes affect differential voltage signals with and without lithium plating \autocite{OKane2020}. The temperature dependence of diffusion is well reproduced by an Arrhenius response \autocite{Ecker2015,Cabanero2018}. While Fickian diffusion is by far the most widely used model of mass transport within particles, there is increasing evidence that other models that account for phase change, such as the Cahn-Hilliard equations \autocite{Cahn1958,Novick-Cohen1984}, may be more appropriate for graphite \autocite{Escalante2020} and LFP \autocite{Bazant2013}, although parameters such as the characteristic transition-region length do not have clear meanings in the context of classical thermodynamics. 

\subsection{\label{section:electrode mft}Measurement/fitting techniques}

\subsubsection{Electronic conductivity probes} \label{sec:eleccondprobe}

Four-point surface probes are commonly used to measure the electronic conductivity of porous electrodes. A current is applied across the sample via two of the probes, while the other two measure the potential drop. The probe spacing is fixed, but varies from device to device: it can range from a few micrometers to a few millimeters \autocite{Jooss2015}. By relating the applied current and the measured voltage, the electronic conductivity can be estimated \autocite{Mandal2001,Chen2007}, using suitable corrections based on sample size and shape \autocite{Smits1958}. Current collectors can also be separated from the electrode coating to reduce their influence on measurements by strategies such as the use of strong adhesives to delaminate the electrode material or solvents to dissolve the current collector foil \autocite{Chen2020,Peterson2014,Illig2012}. Due to the anisotropy of the electrode coatings, conductivity depends on the orientation of the sample \autocite{Tian2019}. In an operating battery, the through-plane conductivity (in the direction perpendicular to the current collectors) is most relevant for use in the DFN model, as this relates to the movement of electrons from the surface of the electrode particles to the current collector \autocite{Kondo2019}. Bespoke methods have also been developed to measure the combined interfacial resistance \autocite{Lanterman2015,Ender2013}.

Through-plane conductivity can be measured using a two-electrode method \autocite{Thorat2011,Bauer2015}. This involves sandwiching the sample between two probes, then evaluating the direct-current resistance. Deconvolution of contact resistances can be a challenge in the two electrode set-up; bespoke four-point probe designs offer greater accuracy \autocite{Wei2016}.

\subsubsection{\label{pOCV} Pseudo open-circuit voltage (pseudo-OCV) test} 

During cycling of an active material, the overall voltage response is comprised of the equilibrium OCV of each electrode plus the overpotentials between each electrode and the electrolyte, and the potential dropped across the electrolyte and solid components of the electrodes. Each of these contributions to the potentials, except the OCVs, generally increase with current throughput. The galvanostatic pseudo-OCV method minimises the non-OCV contributions by sweeping through a cell's stoichiometric range at a slow rate (e.g., less than C/20 --- i.e., a current at which the rated capacity of the cell would charge or discharge in 20 hours) to determine the cell OCV. It should be noted that this method never truly reaches an open-circuit condition, but can be made to come arbitrarily close to it by decreasing the C-rate. \textcite{Chen2020} compared the pseudo-OCV approach to the GITT technique described in section \ref{gitt}, concluding that polarisation, kinetically stable phase transitions and hysteresis are still observed in most pseudo-OCV data.

\subsubsection{\label{gitt} Galvanostatic intermittent titration technique (GITT)}

\begin{figure}
\centering
    \includegraphics[width=0.8\textwidth]{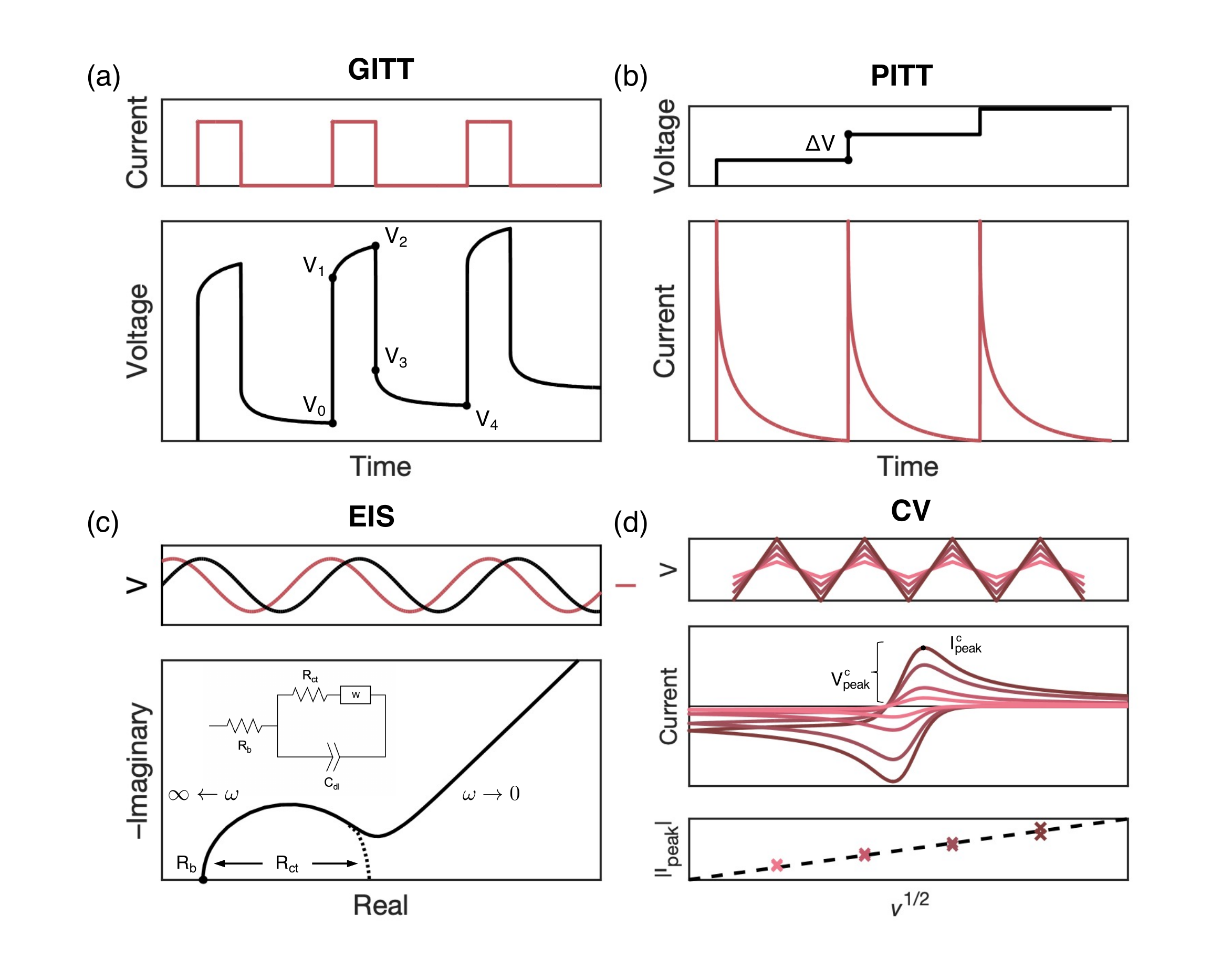}
    \caption{Exemplary signals from characterisation techniques for galvanostatic intermittent titration technique, GITT (a), potentiostatic intermittent titration technique, PITT (b), electrochemical impedance spectroscopy, EIS (c), and cyclic voltammetry, CV (d). In plot (c), an example Nyquist diagram for a Randles circuit (inset) is illustrated.}
    \label{fig:GITT_PITT_EIS_CV}
\end{figure}

\textcite{Weppner1977} introduced the Galvanostatic Intermittent Titration Technique (GITT) in 1977 to characterise diffusivities in dense planar electrodes. The method has since been adapted for porous electrodes. GITT involves applying a short and weak current pulse, followed by an open-circuit (no applied current) rest period which is sufficiently long that a cell voltage close to the equilibrium potential is obtained. The charge transferred should be small enough such that the voltage response remains proportional to the current applied, and data should only be processed outside the timescale dominated by capacitive relaxation of the electrolytic double layer. An exemplary set of GITT data is provided in figure \ref{fig:GITT_PITT_EIS_CV} (a). 

For intercalation electrodes, GITT is used to establish electrode OCVs at different states of lithiation \autocite{Wen1980,Chen2020,Cui2016,Ko2019}. After each current pulse and subsequent relaxation step, the equilibrium OCV, $U_k$, is given (to a good approximation) by the cell voltage, and the lithiation concentration $c_{k0}$ is known from the pulse current and duration through Faraday's law of electrolysis. 
It should be emphasized that thermodynamic equilibrium is only well approximated by GITT-derived OCVs that have been recorded with sufficient rest periods after each pulse.



The quantification of the solid-state diffusivity from GITT data is based on an analysis with Fick's second law. Assuming linear diffusion limited by the solid electrode material, and that the pulse duration is short enough that diffusion within the electrode film can be assumed semi-infinite, the diffusivity $D_k$ of lithium within the solid is given by
\begin{equation}\label{eq:GITT_Ds_general}
    D_k = \frac{4}{\pi} \left( \frac{i_\mathrm{app} A}{S_k F} \right)^2 \left( \frac{U'_k(c_{k0})}{\frac{\partial V}{\partial\sqrt{t}}} \right)^2, 
\end{equation}
where 
$A$ is the geometrical area of the porous electrode, $S_k$ is the area of the electrode-electrolyte interface within it, $c_{k0}$ is the Li$^+$ concentration in the solid at the beginning of the pulse,  $U'_k(c_{k0})$ is the derivative of OCP with respect to concentration at $c_{k0}$,  and $\tfrac{\partial V}{\partial\sqrt{t}}$ is the derivative of the terminal voltage with respect to the square root of time. Note that this analysis assumes a semi-infinite film penetration model of diffusion \autocite{Bard2001}; it is important to be aware of whether assumptions concerning the nature of diffusion are applicable to the system being studied \autocite{Yang2019}. 


The GITT method was originally derived for compact thin-film electrodes \autocite{Weppner1977}, but remains valid for porous electrodes, as long as $S_k$ is taken to be the effective surface area available for charge transfer within the porous material and the configuration is such that spherical diffusion within the electrode particles does not become rate limiting. Generally, $S_k$ relates to the DFN surface-to-volume parameter $b_k$, through 
\begin{equation}\label{eq:surface_area}
    S_k = b_k A L_k ,
\end{equation}
where $L_k$ is the thickness of the porous electrode. Assuming that the electrode comprises spherical particles, all of radius $r_k$, that occupy a volume fraction $\varepsilon_{\mathrm{act},k}$ of the active material, and that the distribution of particles throughout the film depth is uniform then, it follows that $b_k = 3 \varepsilon_\mathrm{act,k}/ r_k$. Application of GITT to porous-electrode models is well detailed by \textcite{Nickol2020}.

A reaction-kinetic overpotential is present during any galvanostatic measurement, hence the GITT set-up should be chosen to minimise any overpotentials observed on the working electrode, maintaining accurate measurement of the thermodynamic states through the OCV.
In two-electrode measurements care should be taken to ensure that these overpotentials are small enough that they do not appreciably contaminated the results. Even when using Li metal counter electrode half cells, the Li metal may contribute an additional overpotential, leading to inaccuracies in the observed cell voltages and underestimation of diffusion properties.  Three- or four-electrode cells, incorporating one or two reference electrodes, produce more accurate voltage responses because they allow the contributions from the overpotential to be quantified and subsequently subtracted off of the data \autocite{Nickol2020}.

For sufficiently small currents and short time intervals, $dV/d \sqrt{t}$ is constant, experimental data can be used to estimate the change in OCV and the charge injected with the pulse directly relates to the concentration change in the solid through Faraday's law of electrolysis. Thus, one can use voltages from the data in figure \ref{fig:GITT_PITT_EIS_CV} (a) to
simplify equation (\ref{eq:GITT_Ds_general}), yielding
\begin{equation}
    D_k = \frac{4}{\pi \tau} \left( \frac{r_k}{3} \right)^2 \left( \frac{V_4 - V_0}{V_2 - V_1} \right),
\end{equation}
 where, apart from the measured voltages $V_j$, the only inputs are the particle radius, $r_k$, and the pulse duration, $\tau$. This approach removes some error associated with overpotentials, since the equilibrium voltage is recorded without an internal resistance contribution. Poor estimates of diffusivity can still occur, however, particularly when GITT is undertaken in an SoC region with a relatively flat OCV response, or if equilibrium is not (approximately) reached by the end of each rest period. An alternative approach is to fit the voltage transient versus $\sqrt{t}$ during the current pulse, and calculate the diffusion coefficient directly using equation \eqref{eq:GITT_Ds_general}. This approach does not eliminate any overpotential errors. More elaborate methods to calculate $S_k$ can be used, as described in section \ref{sec:geomparams}, but the value should always be consistent with that used in the DFN model.



Recently, there have been several developments in the application and analysis of this technique. A modified GITT technique, using a `staircase' current profile, has been reported to determine other kinetic parameters, including the exchange current density and charge transfer coefficients, by fitting the observed overpotentials to a modified Butler–Volmer equation \autocite{Heubner2016}. This `staircase' current profile, applied before the GITT pulse, consists of a set of very short, constant current pulses of different amplitudes, which make no net contribution to the state of charge. High current galvanostatic methods, such as high power pulse characterisation (HPPC), may create large, local surface concentrations which increase the observed polarisation in the voltages, but which can be used to extrapolate the different resistance contributions within the electrode and cell. Other developments in the estimation of diffusion coefficients have focused on the methods to fit the model to the data \autocite{Nickol2020,Shen2013}.

\subsubsection{\label{cvmeth}Cyclic voltammetry}

Cyclic voltammetry (CV) is a technique commonly employed to investigate reduction and oxidation processes \autocite{Bard2001}, using a three electrode system - working, counter and reference electrodes - and measuring the current response to a linearly cycled potential sweep between two given voltage limits, as depicted in Figure \ref{fig:GITT_PITT_EIS_CV} (d). In sweep voltammetry, an applied potential is varied linearly with time while recording the resultant current. This potential sweep is characterised by a particular scan rate, $\nu$ (with units V/s). Data is typically reported in the form of a cyclic voltammogram, which plots current versus voltage, both parametric functions of time. Analysis of cyclic voltammograms provides both qualitative and quantitative information about electrode reaction kinetics and mass transport. For electrode parameterisation, CV can be used to obtain kinetic rate constants of (de)lithiation \autocite{Levi1999,Vassiliev2016} and the diffusion coefficient of lithium within electrodes \autocite{Benedek2020}. These techniques are discussed in detail in \textcite{Kim2020}.

The quantitative analysis of cyclic voltammograms varies according to whether the electrochemical reaction is classified as reversible (i.e. kinetically fast), quasi-reversible, or irreversible (i.e. kinetically sluggish). For a reversible reaction, the forward and reverse electron-transfer steps are in equilibrium and voltage follows a Nernst-like equation. Li-ion intercalation is often considered to be reversible \autocite{Pender2020}, though quasi-reversible analyses have also been applied when rates of charge transfer and mass transport are similar \autocite{Yu_2007}.

A reversible, single-electron transfer cyclic voltammogram has the following characteristics: a Faradaic response is observed, meaning that the difference in the oxidative and reductive peaks is ~56.5 mV at room temperature, the ratio of the amplitude of the current peaks is unity ($|I^\textrm{a}_\textrm{peak}/I^\textrm{c}_\textrm{peak}|=1$) and, given a diffusion-controlled process, the peak current is proportional to the square root of the scan rate ($I_\textrm{peak} \propto \sqrt{\nu}$). 

For porous electrodes, the surface area and morphology of the surface also affects the shape of the voltammogram. As many common models that underpin cyclic voltammetry analysis are derived based on assumptions that the working electrode is small compared to the counter and that liquid-phase diffusion is semi-infinite, translation of the theoretical equations to represent porous electrodes is difficult. The effects on peak voltage and reversibility are particularly pronounced on thicker porous electrodes \autocite{Yu_2007}. The impacts of intercalation-particle geometry and dispersion on CV profiles have also been modelled \autocite{Vassiliev2016}. It was found that accounting for the real particle-size distribution may be crucial for determining the voltammetric response. When mass-transfer contributions derive from both free electrolyte adjacent to the electrode and pore-filling electrolyte within it, reversible reactions may also produce an apparent shift in peak voltage position, proportional to the scan rate \autocite{Yu_2007}.

In general, thin electrode coatings minimise the observed kinetic limitations and the effect of the porous electrode geometry. Given a reversible CV response, the relationship between the peak current, $I_\textrm{peak}$, and the scan rate, $\nu$, see figure \ref{fig:GITT_PITT_EIS_CV} (c), can be fit with the following equation to elucidate the solid-phase diffusivity of lithium, $D_k$ \autocite{Yu_2007,Takahashi2002,Bard2001}:
\begin{equation}
I_\textrm{peak} = 0.4463 F c_k S_k \sqrt{\frac{F \nu D_k}{RT}}.
\end{equation}
Here $c_k$ is the concentration on the surface of the particle and $S_k$ is the effective surface area per unit mass. The effective surface area can be further refined to account for the electrochemically active surface area. Since practical battery electrode materials more often exhibit quasi-reversible or irreversible characteristics, CV parameterisation techniques may only offer qualitative comparisons between effective kinetic and transport properties.

\subsubsection{\label{potentio}Potentiostatic methods}
Potentiostatic methods involve applying a voltage step and observing the current response. Examples include a voltage sweep, such as in CV, or constant voltage measurements, such as chronoamperometry. Chronoamperometry can also be applied over a range of concentrations or states of charge of the electroactive material and, at each concentration (or voltage), the current transient can be used to calculate the solid-state diffusion coefficient. 

In the potentiostatic intermittent titration technique (PITT), a stepped constant voltage is applied and the current decay is tracked, at each voltage, to a set current limit. A rest period can also be introduced in between each voltage step, after the current decays such that open-circuit voltage is (approximately) reached \autocite{Weppner1978,Wen1979}. The exclusion of rest periods accelerates data collection. However, including the rest period in PITT means the titration method can be used to calculate the diffusion coefficients while minimising overpotentials that contribute errors to the concentration estimation. Care must be taken with any overpotential observed in the measurement and, again, 3-electrode measurements with a reference electrode are significantly more accurate. A typical PITT data set is depicted schematically in figure \ref{fig:GITT_PITT_EIS_CV} (b). 

A model based on Fickian diffusion is used to estimate diffusivity from current data \autocite{Wen1979}. In the case where $t \ll {r_k^2}/{D_k}$ (short times compared to the time scale of solid-state diffusion), and where interfacial reaction kinetics is reversible, it is appropriate to apply the Cottrell equation \autocite{Bard2001}:
\begin{equation}\label{eq:PITT_Cottrell}
    D_k = \pi t \left(\frac{i(t)}{FS_k(c_{k\mathrm{s}}-c_{k0})}\right)^2,
\end{equation}
whereas for $t \gg {r_k^2}/{D_k}$ (long times) the current decay becomes exponential and the solid-state Li diffusion coefficient can be determined from \autocite{Wen1979,Xie2009}:
\begin{equation}\label{eq:Dexp}
    D_k = -\frac{4 r_k^2}{\pi^2} \frac{\partial \ln i}{\partial t} .
\end{equation}
 This has been expressed in terms of a particle radius, $r_k$, and, within the caveats described in section \ref{gitt}, it is true to say that $b_k = 3 \varepsilon_\mathrm{act,k}/ r_k$.

\textcite{Malifarge2017} have provided guidelines for the implementation of PITT and its underlying assumptions. Less conventionally, PITT can also evaluate the exchange current density \autocite{Li2012} but, as already mentioned (section \ref{sec: EIS Geometry}), EIS is more often used. Diffusion coefficients via PITT and GITT have been shown to corroborate one another \autocite{Han2004,Levi2012}. The potentiostatic methods offer a speed advantage over galvanostatic methods, but additional measures may be needed to minimise the potential difference between the working electrode and the reference or counter electrode, and the voltage drop across the electrolyte. Not properly accounting for these overpotentials can lead to overestimates of the solid-phase diffusion coefficients.

\subsubsection{\label{frequency}  Electrochemical impedance spectroscopy}

The EIS technique was described in section \ref{sec: EIS Geometry} of this review and is revisited again in section \ref{sec:elyteEIS}. In the context of electrode parameterisation, EIS may be used to estimate charge-transfer resistance, double-layer capacitance and diffusion timescales, usually by fitting parameters of a simple circuit model to EIS data, by means described below. A typical model is the Randles circuit which, per electrode, consists of a double layer-capacitance in parallel with the series combination of a Warburg element, representing diffusion, and a charge-transfer resistance, as shown in figure \ref{fig:GITT_PITT_EIS_CV} (c). Further series resistances may be added to represent additional Ohmic and ionic voltage drops, as well as additional elements to represent electrolyte dynamics, if required.


\textbf{Charge Transfer.}
An expression for charge-transfer resistance may be derived by linearising the Butler-Volmer equation (\ref{fc5}) for small values of the overpotential, \ie about $\eta_k=0$. On doing so, one finds that the charge transfer resistance (per electrode), $R_\text{ct}$, relates to exchange current density through:
\begin{equation}\label{eq:CT_resistance}
    j_{k0} = \frac{RT}{S_k FR_\text{ct}} = \frac{RT}{A L_k b_k FR_\text{ct}}.
\end{equation}
Measurements of $R_\text{ct}$ by EIS have thus been used to determine kinetic parameters \autocite{Doyle2000,Swiderska-Mocek2017}.

%

Large variances among calculated exchange current densities $j_{k0}$ stem, in part, from the choice of surface area, with different groups using different definitions, as described by \textcite{Chang2000} and \textcite{Verbrugge1999}. The exchange current density relates to the Butler--Volmer rate constant, $K_k$, through equation \eqref{fc7}.

The identification of the $R_\text{ct}$ value associated with the main intercalation reaction for individual electrodes is a challenging task. Without a reference electrode, it is not possible to distinguish between the separate half-cell reactions (recall that in the model, there are in fact two Butler–Volmer equations, one for each electrode). Another issue is that the impedance response of the active materials and the response associated with the constantly evolving passivation surface films tend to occur in similar frequency ranges \autocite{Zhuang2007}. Changing porosity and surface roughness also leads to frequency dispersion \autocite{Pistoia1996}. The majority of strategies to fit $R_\text{ct}$ from overall spectra involve performing EIS measurements under conditions that cause the kinetic and surface-layer behaviour to diverge. This includes observing dependence on temperature \autocite{Chen2020}, changes with conditioning cycles to develop SEI \autocite{Swamy2015}, or varying SoC \autocite{Amin2017}. Insight can also be gained by comparing EIS spectra obtained from full-cell and half-cell configurations \autocite{Schmalstieg2018}. 
The temperature dependence of $R_\text{ct}$ has been modelled in terms of the activation energy for lithium transport through the SEI layer \autocite{Keefe2019}.

\textbf{Diffusion.}
While charge transfer occurs at medium to high frequencies, the low-frequency EIS response of a porous electrode provides information about the solid-state diffusion coefficient in each electrode \autocite{Sikha2007}. The considerations mentioned above also apply here, namely that, without a reference electrode, it is not possible to distinguish between the separate electrodes (note that this issue is not unique to EIS, it also applies to GITT and other techniques mentioned in this section). In the frequency domain, diffusion transport is characterised by a Warburg element, which may be derived for various different geometric configurations, including thin films, as well as planar, cylindrical, or spherical particles (see \textcite{Song_2012}). For spherical particles, the diffusion-related impedance for each electrode $k$ may be expressed as \autocite{Bizeray2019}:
\begin{equation}
	Z_{\text{diff,sp,k}} =  U'_k(c_{k0}) c_k^{\max}  \frac{\tau_k^d}{3 Q_k}  \frac{ \tanh{\left(\sqrt{s \tau_k^d}\right)} }{ \tanh{\left(\sqrt{s \tau_k^d}\right)} - \sqrt{s \tau_k^d} } .
\end{equation}
where $s$ is the Laplace variable, $U'_k(c_{k0})$ is the gradient of half cell OCV, with respect to concentration, $\tau_k^d$ is a diffusion timescale ${r_k^2}/{D_k}$ for each electrode, and $Q_k = F A L_k \varepsilon_{\textrm{act},k} c^{\max}_{k}$ is the maximum capacity of each electrode.
%

\textcite{Ecker2015} demonstrated good agreement between the diffusion values obtained by GITT and EIS. Further comparison of the techniques for characterising diffusion coefficients can be found in \textcite{Deng2020a} and \textcite{Ivanishchev2017}. Applying titration techniques for SoC-dependent diffusivities is also particularly time-consuming (ca.\ 17 days \autocite{Barai2019}) and EIS may be a more time-efficient alternative. Similar to most parameters with Arrhenius-like responses, the activation energy can be determined by fitting repeat measurements at different temperatures \autocite{Ecker2015,Schmalstieg2018}.


\section{Electrolyte parameters}
\label{sec:elyteparams}


Intrinsic electrolyte properties that parameterise DFN-type models are primarily concerned with how ionic and neutral species in solution are redistributed under the influence of electric fields, concentration (electrochemical potential) gradients and bulk convection. Transport properties are sensitive to salt and solvent composition, as well as temperature.

Ion transport within the electrolyte for DFN models was originally described using multi-component concentrated solution theory  \autocite{NewmanElectrochemical,doyle1993modeling,Chapman1973,GLi2020}. As such, the three familiar transport properties of ionic conductivity, $\sigma_\mre$, salt diffusivity, $D_\mre$, and transference number, $\tp$, are related to a set of three pairwise Stefan--Maxwell diffusivities, which account for frictional drag between solvent and cation, solvent and anion, and cation and anion, respectively. Dilute-solution theory has also been applied in the literature, to model the electrolyte phase in batteries \autocite{Ecker2015}. Key assumptions made for this approach include non-interacting ions, infinitely dilute salt concentration and thermodynamic ideality ($1+\partial\ln f_{\pm}/\partial\ln c_{\textrm{e}} = 1$). \textcite{Bizeray2016}, as well as \textcite{Richardson2020}, review the discrepancies between dilute, moderately-concentrated, and concentrated models. In typical battery applications, concentrated electrolytes are used, and concentrated solution theory is better suited to describe the underlying physics, as well as to model a broad range of scenarios such as fast charging or complex multi-component electrolytes.




\subsection{Parameters}

\subsubsection{Ionic conductivity}

\begin{figure}
    \centering
    \includegraphics[width=0.7\textwidth]{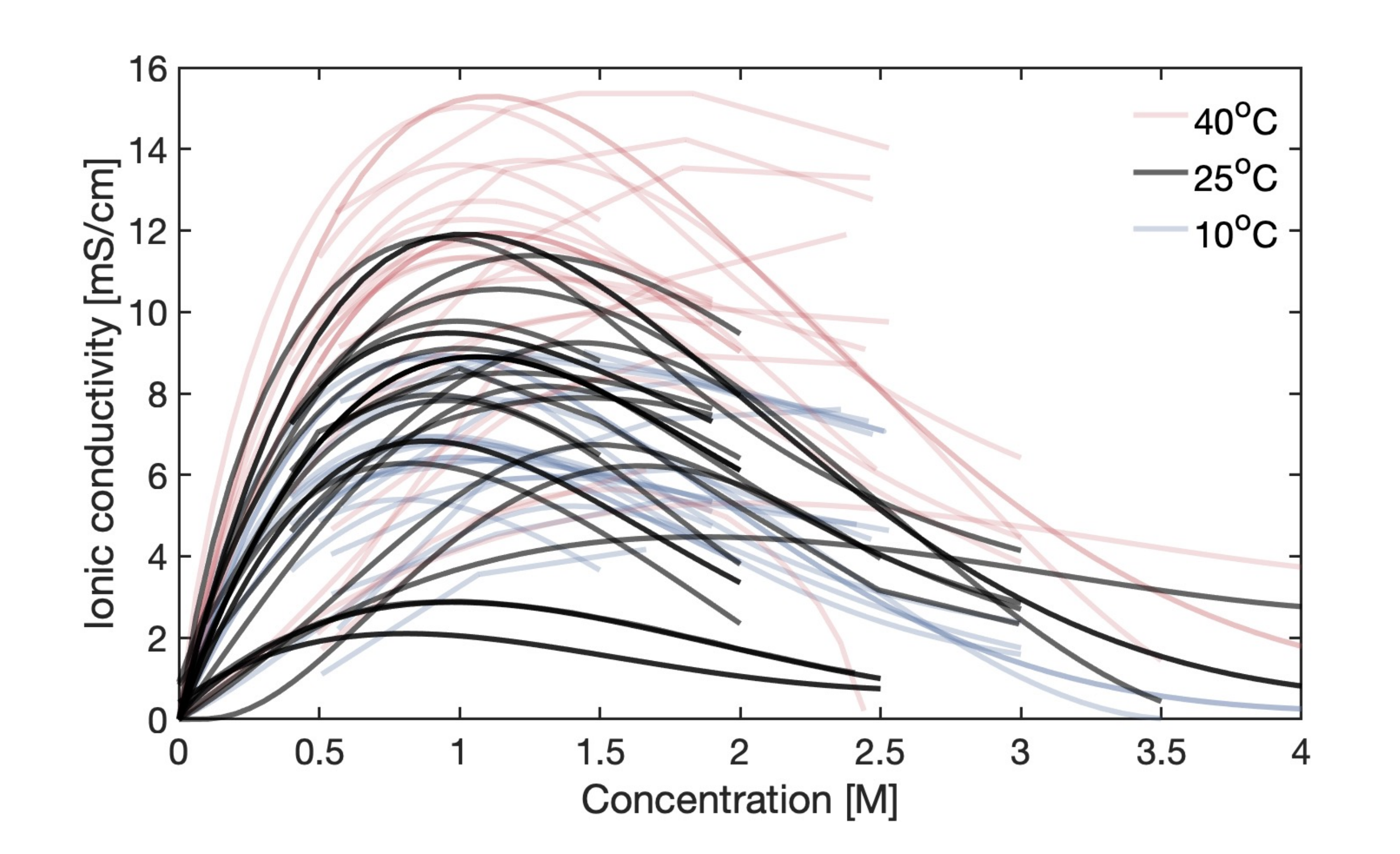}
    \caption{Ionic conductivities of LiPF$_6$-containing electrolytes, at temperatures of 10 $^{\circ}$C (blue), 25 $^{\circ}$C (black) and 40 $^{\circ}$C (red), reported from the LiionDB literature parameter database \autocite{LiionDB}.}
    \label{fig:elytecond}
\end{figure}

The ionic conductivity, $\sigma_\mre$, quantifies the mobility of ions under an electric field, in the absence of concentration gradients. It appears in the generalised Ohm's law (equation \ref{fc2}) and determines the  ionic current in response to the gradient of the electrochemical potential. Ionic conductivity is a strong function of temperature and concentration. In dilute electrolytes, conductivity rises with salt concentration, as the number of mobile ions in solution increases. As salt content increases past a certain threshold, however, species--species interactions, such as ion pairing, cause $\sigma_\mre$ to fall \autocite{Ding2001}. The dependence of conductivity on salt concentration at a given temperature can be fit well with simple polynomial forms \autocite{Nyman2008} or by phenomenologically inspired forms, such as the square-root dependence of equivalent conductance ($\sigma_\mre/(Fc_\mre)$), predicted by Debye--H\"{u}ckel limiting behaviour \autocite{Debye1923,Onsager1932,Hou2020}. Other empirical functional forms have also been employed, to account for temperature dependence \autocite{Valoen2005,Landesfeind2019}. 

Ionic conductivities, $\sigma_\mre$, for a variety of LiPF$_6$-containing electrolytes from the property database are presented in figure \ref{fig:elytecond}. Generally, ion transport improves with temperature. The temperature dependence of electrolyte conductivity in practical conditions is well-described by a simple Arrhenius relation \autocite{Ding2001,Ecker2015}. Consistent results may be attained by interpolating conductivities using activation energies fit from a range of temperatures. Extrapolating to wider temperature ranges may require compensating for thermal effects on the dielectric constant and viscosity \autocite{Petrowsky2009,Xu2004}.



In LiBs, the electrolytic solution permeates porous phases in the separator and electrodes. Therefore, before they can be utilised in the DFN model, effective transport parameters must be obtained by scaling with a transport efficiency factor, $\mathcal{B}$. These extrinsic details were discussed in section \ref{sec:porephasetransport}. 

\subsubsection{Salt diffusion coefficient}
\label{sec:diff}

\begin{figure}
    \centering
    \includegraphics[width=0.7\textwidth]{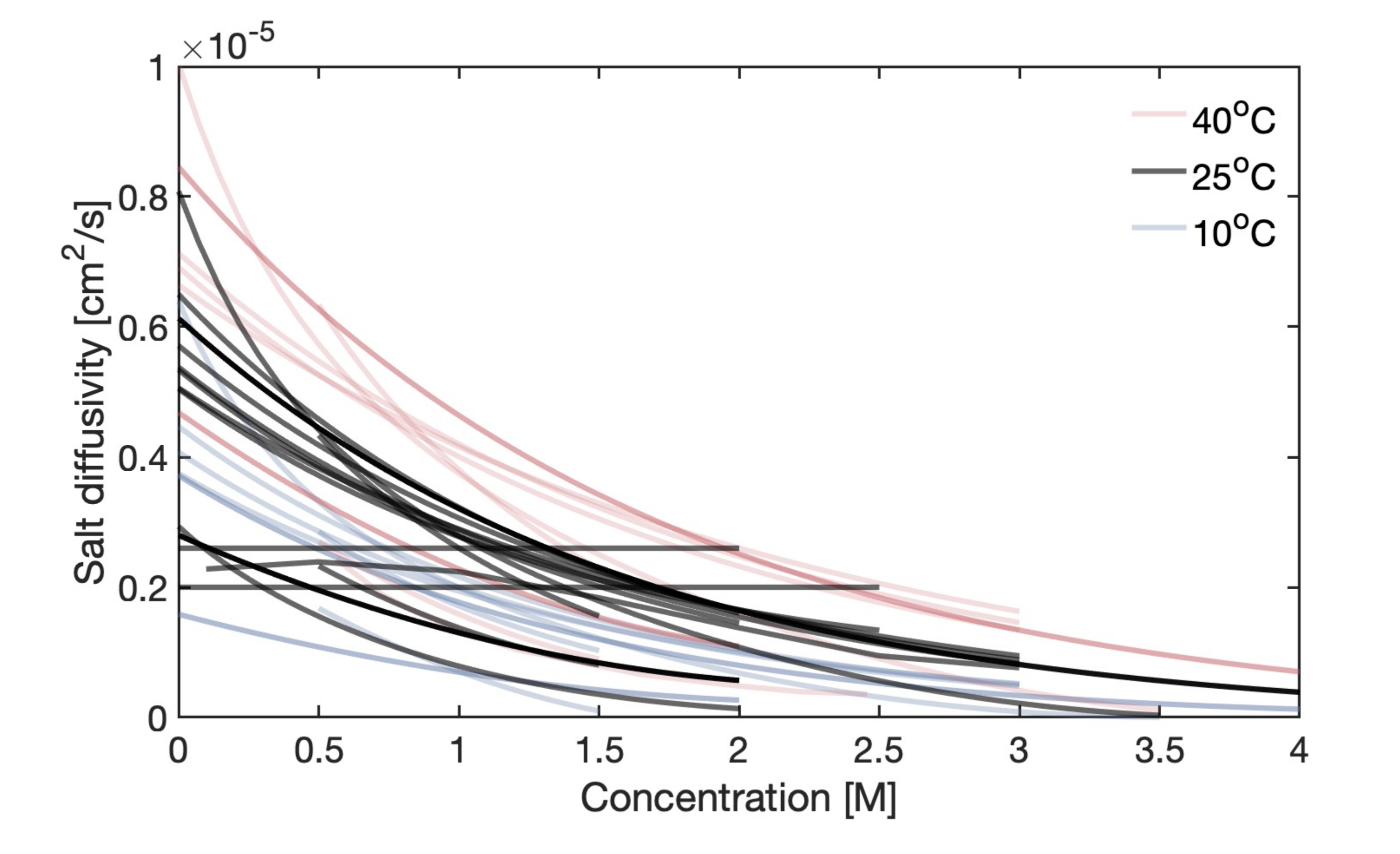}
    \caption{Salt diffusivities of LiPF$_6$-containing electrolytes, at temperatures of 10 $^{\circ}$C (blue), 25 $^{\circ}$C (black) and 40 $^{\circ}$C (red), reported from the LiionDB literature parameter database \autocite{LiionDB}.}
    \label{fig:elytediff}
\end{figure}

The salt diffusivity, $D_\mre$, determines the contribution to the ionic flux by salt concentration gradients, as given by equation \ref{fc1}. The thermodynamic diffusivity, $\mathscr{D}$, also describes diffusion under the true driving force of chemical potential gradients \autocite{Chapman1973}. A direct conversion between $\mathscr{D}$ and $D_\mre$ --- the value usually experimentally measured --- exists via the thermodynamic factor, $1+\partial\ln f_{\pm}/\partial\ln c_{\textrm{e}}$ (discussed below in section \ref{sec:tdf}):
\begin{equation} 
    \label{eq:thermdiff}
    D_\mre = c_\text{T}\overline{V}_0\left(\tdf\right)\mathscr{D}.
\end{equation}
The additional term, $c_\text{T}\overline{V}_0$, which is the total molarity, $c_\text{T}$, multiplied by the solvent partial molar volume, $\overline{V}_0$, arises as a drift factor that puts $D_\mre$ in reference to the solvent velocity. It is this diffusion coefficient, $D_\mre$, which most commonly appears in DFN model equation \eqref{fc1}. Diffusion coefficients have been demonstrated to decrease with increasing salt concentration and solvent viscosity, as measured by a range of techniques. The rate of diffusion also increases with temperature. Figure \ref{fig:elytediff} shows the salt diffusivities, $D_\mre$, reported in the literature across varying solvent compositions and salt concentrations. Diffusion coefficients are typically well fitted by simple polynomial expressions or exponentials. 

Salt diffusion coefficients can be measured experimentally (see restricted diffusion below, in subsection \ref{sec:polarisation-cells}, \autocite{Hou2020,Wang2020}). Models that use a constant, average diffusivity will inherently force concentration gradients in solution to develop symmetrically \autocite{Prada2012}. This may cause an over- or underestimation of the concentration polarisation. Many modeling efforts use approximate diffusion coefficients. \textcite{Schmalstieg2018} use the Nernst--Einstein relationship, derived from ideal, infinitely dilute assumptions, to estimate $D_\mre$ from $\sigma_\mre$. The Nernst-Einstein model is not well justified for concentrated solutions, however, and deviations from this ideal behaviour occur well below the salt concentrations of around 1 M that are typical in LiB electrolytes \autocite{Bizeray2016}. Self-diffusivities attained by nuclear magnetic resonance (NMR) or molecular dynamics (MD), while similar in magnitude to $D_\mre$, require additional interpretation to be used consistently in the DFN model \autocite{Kim2016, Fong2019}.


The choice of model used to extract the diffusion coefficient from experimental data strongly influences the inferred diffusivity. For example, following  \textcite{fuller1994simulation}, the ion flux (equation \ref{fc1} in our macroscopic relationships) is composed of contributions from migration and diffusion, with convection assumed to be negligible. The majority of DFN battery models have adopted this convention. Indeed, if convection, driven by the density changes which accompany concentration changes, is accounted for, as discussed by the works of Liu and Monroe \autocite{Liu2014,Liu2015}, this may skew the apparent diffusion coefficient and transference number.

\subsubsection{Transference number}
\label{sec:transference}

\begin{figure}
    \centering
    \includegraphics[width=0.7\textwidth]{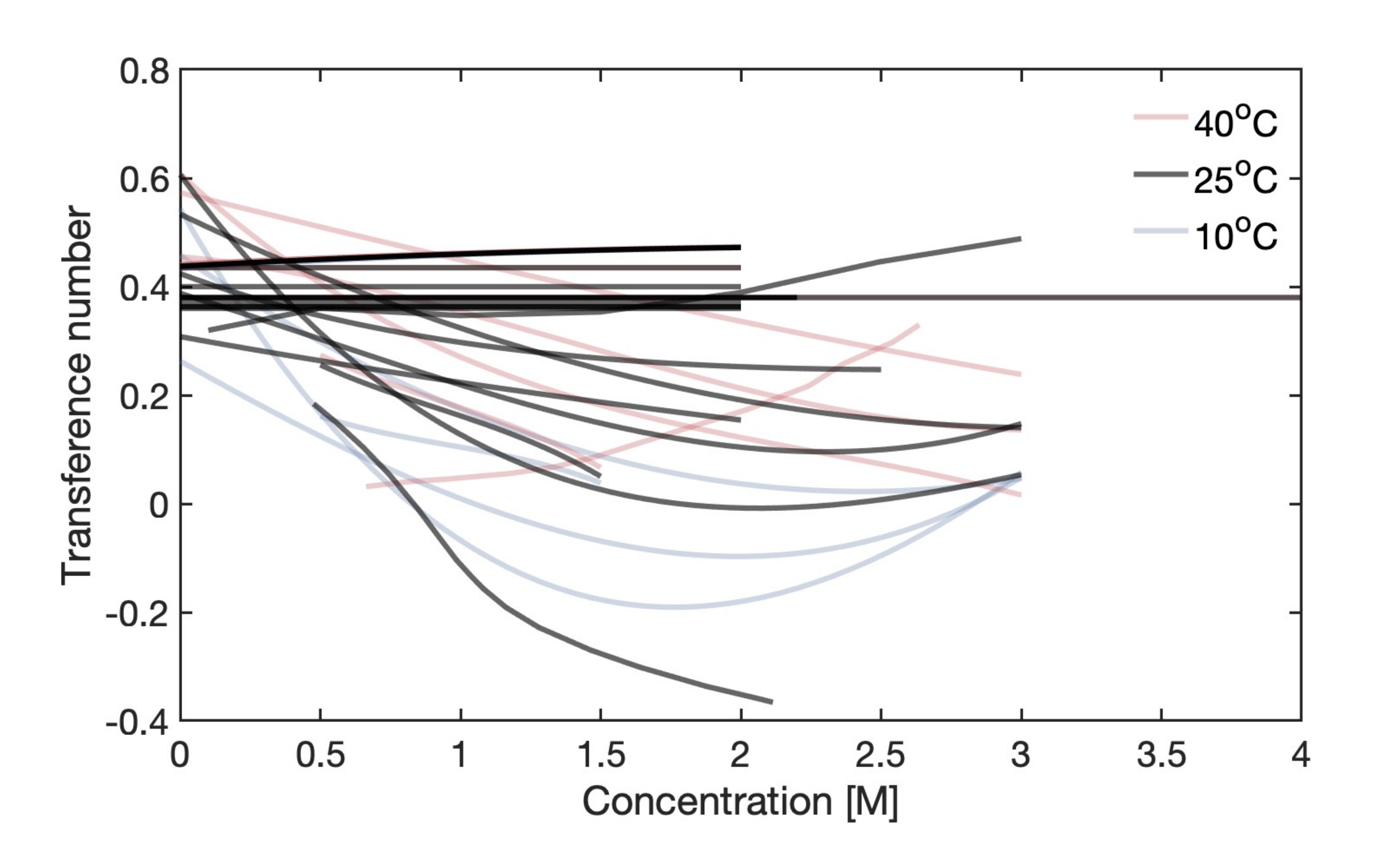}
    \caption{Transference numbers of LiPF$_6$-containing electrolytes at temperatures of 10 $^{\circ}$C (blue), 25 $^{\circ}$C (black) and 40 $^{\circ}$C (red), reported from the LiionDB literature parameter database \autocite{LiionDB}.}
    \label{fig:elytetrans}
\end{figure}

The transference number, $\tp$, is the fraction of the conductivity contributed by cation motion. This parameter is typically reported relative to a reference frame set by the solvent velocity, which is taken to be zero in conventional DFN models. Cation and anion transference number must sum to unity. From figure \ref{fig:elytetrans} it is clear that the lithium ion carries a minority of the charge during typical battery operation. Many battery models choose to use a simplified average $\tp$ value in the 0.3--0.4 range.

A low transference number implies poor power efficiency. \textcite{Diederichsen2017} demonstrated, via a sensitivity analysis on the DFN model, that raising the transference number by 0.2 can have significant impact on accessible capacity during charge and discharge. The limiting current --- the maximum current at which charge carriers are completely depleted at an electrode --- is also a strong function of transference and diffusivity. 

While it is clear that accuracy of the transference number has a strong impact on the accuracy of battery models at high C-rate, consensus is low on the expected magnitudes and composition dependencies of $\tp$. The varied experimental approaches and processing techniques, using dilute or concentrated solution theory, as discussed in section \ref{section:elyte mft}, also contribute to the spread in the data across all electrolyte formulations and salt compositions \autocite{Zugmann2011}. For example, experimentally obtained negative cation transference numbers have been rationalised by the presence of charged or neutral aggregates \autocite{Villaluenga2018}, or by moving boundaries due to lithium plating \autocite{Richardson2018}. The choice of reference velocity (e.g., solvent or volume-averaged) used can also impact apparent transference \autocite{Chapman1973,Liu2014}.

The ultimate purpose of these electrolyte parameters is to predict the lithium ion concentration, as a function of position and time. It is important to reiterate that the specific formulation of a DFN model will impact the interpretation of experiments designed to measure $\tp$. For example, models that incorporate dendrite growth \autocite{Sethurajan2019} will predict vastly different transference numbers to models with ion pairing \autocite{Richardson2018}. Similarly, models that do not include solute-volume effects \autocite{Liu2014} will overpredict transference when convective-polarisation suppression is wrongly accounted for as a migration effect \autocite{Hou2020,Wang2020}. Finally, it should be borne in mind that, at low C-rates, simplifying assumptions about $\tp$ have minimal impact on simulation validity, but that these assumptions may be violated at higher C-rates. 

\subsubsection{Thermodynamic factor}
\label{sec:tdf}

\begin{figure}
    \centering
    \includegraphics[width=0.7\textwidth]{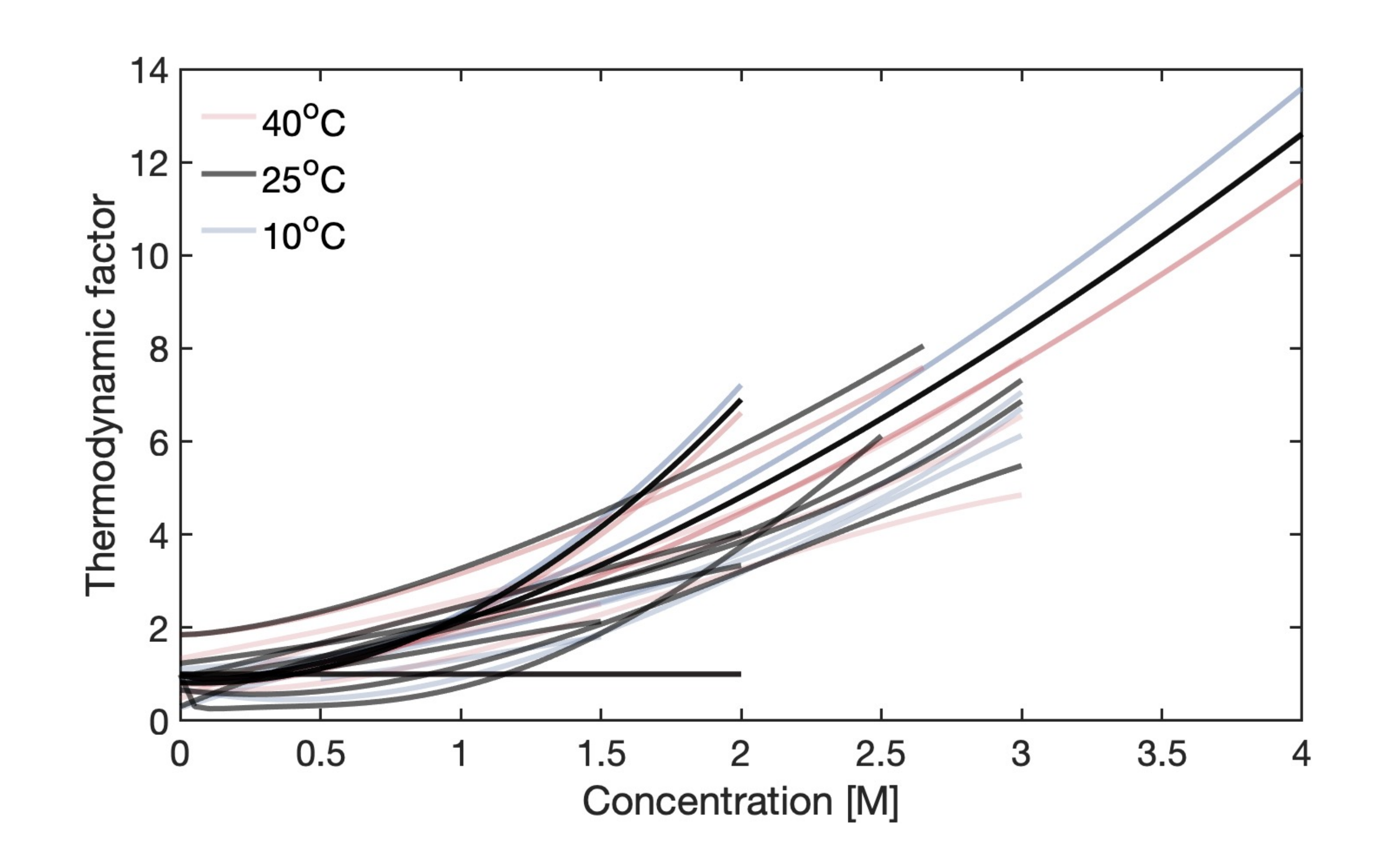}
    \caption{Thermodynamic factors of LiPF$_6$-containing electrolytes at temperatures of 10 $^{\circ}$C (blue), 25 $^{\circ}$C (black) and 40 $^{\circ}$C (red), reported from the LiionDB literature parameter database \autocite{LiionDB}.}
    \label{fig:elytetdf}
\end{figure}

The thermodynamic factor, $(1+\partial\ln f_{\pm}/\partial\ln c_{\textrm{e}})$, is key to translating a concentration difference into an OCV drop, as can be seen from the generalised form of Ohm's law in equation \eqref{fc2}. Thermodynamic factors measure solution non-ideality, accounting for deviations from Nernstian behaviour. The thermodynamic factor is a grouping of terms which arises when the derivative of the chemical potential, $\mu_\mre$, with respect to concentration, is computed \autocite{NewmanElectrochemical}. This leads to the appearance of the derivative of the mean molar salt activity coefficient, $f_{\pm}$, with respect to salt molarity, $c_\mre$.  In the literature, the gradient of the activity coefficient is also often defined relative to composition in a molal ($m_\mre$) basis, which can be mapped from the molar basis via the mean molal activity coefficient, $\gamma_\pm$, solvent molarity, $c_0$, and partial molar volume of solvent, $\overline{V}_0$:
\begin{equation}
    \label{eq:tdf}
    \tdf = \frac{1}{c_0\overline{V}_0}\left(1+\frac{\partial \ln{\gamma_\pm}}{\partial \ln{m}}\right)= \frac{1}{RT}\frac{\partial \mu_\mre}{\partial \ln{c_\mre}}.
\end{equation}
By definition, the thermodynamic factor approaches unity at infinite dilution. Debye-H\"uckel theory predicts a decrease in thermodynamic factor with respect to concentration near infinite dilution \autocite{Debye1923,Onsager1932, RobinsonStokes1959}, but this concentration range is far from the practical concentrations used in lithium-ion electrolytes. The LiB literature, which focuses on moderately dilute to concentrated electrolytes, typically reports $1+\partial\ln f_{\pm}/\partial\ln c_{\textrm{e}}$ values above unity that increase across the observed concentration range, in agreement with a condition where ion--solvent or ion--ion interactions are significant \autocite{Wang2020}. There also appears to be a weaker dependence on temperature than for other parameters \autocite{Landesfeind2019}. Polynomials or power series in $\sqrt{c}$, following Debye-H\"{u}ckel theory, have been used to fit correlations for $1+\partial\ln f_{\pm}/\partial\ln c_{\textrm{e}}$ \autocite{Hou2020,Wang2020}.

The measurements summarised in figure \ref{fig:elytetdf} indicate that conventional lithium-ion electrolytes with compositions above 0.5 M should not be treated as ideal \autocite{Stewart2008,Wang2020,Landesfeind2019}. The original DFN publications included $1+\partial\ln f_{\pm}/\partial\ln c_{\textrm{e}}$ in their model formulation, but assumed ideality because thermodynamic data was unavailable \autocite{doyle1993modeling,fuller1994simulation}. Many subsequent DFN models for conventional LiBs have dropped $1+\partial\ln f_{\pm}/\partial\ln c_{\textrm{e}}$ \autocite{Ecker2015,Schmalstieg2018,Chen2020}.
Inspection of the DFN model formulation reveals that neglect of the thermodynamic factor has no effect on compositions predicted by the diffusion equation in equation \ref{fc1}, but exclusion of $1+\partial\ln f_{\pm}/\partial\ln c_{\textrm{e}}$ will introduce errors in the concentration overpotential, as seen in equation \ref{fc2}. In modelling, this will impact predicted battery capacity, due to the propensity to underpredict cell-voltage cutoffs during charge/discharge. Ideal assumptions only have a strong influence at medium-high C-rates, however, when Ohmic losses are not dominant. In situations where battery operation induces a high degree of concentration polarisation, or for cell formulations with particularly high electrolyte molarities, inclusion of an accurate $1+\partial\ln f_{\pm}/\partial\ln c_{\textrm{e}}$ is warranted. 

\subsubsection{Partial molar volumes} \label{sec: PMV}
Partial molar volumes are thermodynamic properties that describe how volume is distributed across the components in solution ( $\overline{V}_\mre$ for salt and $\overline{V}_0$ for solvent). \textcite{doyle1993modeling} simplified their treatment of concentrated solution theory by ascribing zero molar volume to the dissolved salt in solution. Partial molar volumes have since been included in DFN-type models \autocite{Chandrasekaran2011,Nyman2008,Liu2014,Wang2020}. 

In the one-dimensional ion conservation equation (Eq. \ref{fc1}), the divergence of flux $\frac{\partial N_\textrm{e}}{\partial x}$ yields an excluded-volume factor ($1-\frac{d\ln c_0}{d\ln c_\mre}$). As solvent molarity $c_0$ is typically a weak function of salt content, $c_\mre$, in typical lithium-ion electrolytes, \textcite{NewmanElectrochemical} use this assumption to neglect the excluded-volume factor, hence setting $\overline{V}_\mre$ to zero. Partial molar volumes relate to the excluded-volume factor  through the solvent concentration, $c_0$, and solution density, $\rho$ \autocite{NewmanElectrochemical}:

\begin{equation} 
    \label{eq:excludedvolume}
  1-\frac{d\ln c_0}{d\ln c_\mre} = \frac{\rho}{c_0 M_0}\left(1-\frac{d\ln\rho}{d\ln c_\mre}\right) = \frac{1}{c_0\overline{V}_0},
\end{equation}

\noindent with $M_0$ being the solvent molar mass. One can measure $\overline{V}_\mre$ and $\overline{V}_0$ readily with densitometry, as discussed in section \ref{sec:densitometry}. Recent spectroscopic techniques have also allowed for the probing of solvent concentrations \autocite{Bazak2020,Fawdon2021}.


\subsection{Measurement/fitting techniques} \label{section:elyte mft}

\subsubsection{\label{sec:elyteEIS}Electrochemical impedance spectroscopy (EIS)}
We have previously discussed the use of EIS in inferring geometrical and electrode intrinsic parameters in sections \ref{sec: EIS Geometry} and \ref{frequency}, respectively. Here, we cover its use for elucidating electrolyte intrinsic properties. Commercially-available conductivity probes which operate on frequency-based principles are commonly used to measure ionic conductivities for many common lithium-ion electrolyte formulations \autocite{Ding2001,Nyman2008,Valoen2005,Lundgren2014,Logan2018,Xiong2018,Wang2020}.

EIS is also performed with symmetric planar cells with blocking (stainless steel) or non-blocking (lithium) electrodes. The ionic conductivity, $\sigma_\mre$, may be extracted by measuring the high-frequency resistance value, $R_\text{b}$, in the bulk electrolyte, relative to the EIS cell geometry, and then leveraging the equation:
\begin{equation}
    \label{eq:elyteEIScond}
    \sigma_\mre = \frac{L_\textrm{s}}{R_\text{b}A},
\end{equation}
\noindent where $L_\textrm{s}$ is the planar inter-electrode spacing and $A$ is the geometrical electrode area. This is often done with an electrolyte soaked porous separator between electrodes and can be used to calculate the effective conductivity, based on the high frequency bulk resistance intercept on a Nyquist plot \autocite{Wohde2016,Hou2020,Shah2017}. The low-frequency impedance response can also determine contributions from the ionic diffusion coefficient, $D_\mre(c_\mre)$, and transference number, $\tp(c_\mre)$ \autocite{Wohde2016,Pollard1989}. This approach is akin to a frequency-domain analysis of repeated polarisation-relaxation experiments, discussed in the following section.


\subsubsection{\label{sec:polarisation-cells}Polarisation-relaxation cells }
Polarisation-relaxation cells refer to symmetric planar lithium metal electrode cells, used to characterise electrolyte properties by performing a sequence of steps where a current or voltage (or lack thereof) is applied. The concentration gradients resulting from polarisation-relaxation steps are then tracked, often being inferred through concentration overpotentials (also called diffusion potentials), which scale with the concentration gradient \autocite{Bizeray2016}. The next few sections outline common polarisation-relaxation experimental techniques used for electrolyte parameterisation.\\

\textbf{Restricted diffusion.} 
The electrolyte salt diffusion coefficient, $D_\mre(c_\mre)$, is best measured with a restricted-diffusion experiment in a sealed vertical cell, where the ion flux directly opposes the direction of gravity. Orientation is important, especially when no porous separator is employed to prevent natural convection \autocite{NewmanElectrochemical} or buoyancy effects \autocite{Klett2012} from marring the method. During restricted diffusion, a nonuniform concentration profile is induced across an electrolyte, often galvanostatically or potentiostatically, before monitoring the relaxation of the concentration difference between two fixed points. As concentration gradients relate to the OCV via equation \ref{fc2}, the diffusion coefficient can be inferred from the exponential relaxation of the cell voltage \autocite{Chapman1973,Valoen2005,Hou2020,Wang2020,Landesfeind2019,Bergstrom2021}:

\begin{equation}
   \ln V_\textrm{OCV} \propto \Delta c_\mre = -\frac{\pi^2 D_\mre^\text{effective}}{L_\textrm{s}^2} \cdot t + \text{const}.
\end{equation}

\noindent where $V_\textrm{OCV}$ is the restricted diffusion cell voltage measured during the open-circuit relaxation, $\Delta c_\mre$ is the difference in concentration between the points where $V_\textrm{OCV}$ is measured, $L_\textrm{s}$ is the separator or bulk electrolyte thickness and $t$ is time. A formal analysis by \textcite{Chapman1973} showed that this linear time dependence of the logarithm of voltage can be used to process data accurately, even when diffusivity varies with respect to concentration. Thus, when concentration differences are sufficiently small, Fickian diffusion with a constant diffusion coefficient can be assumed and the diffusivity resulting from a restricted-diffusion measurement should be interpreted as the value at the electrolyte's equilibrium concentration.

\textcite{Ehrl2017} found that the restricted-diffusion method through steady-state polarisation and long-term relaxation was less susceptible to the influence of double-layer relaxation effects than pulse-polarisation methods. Error due to surface effects may also be avoided by using a four-electrode cell, containing two reference electrodes positioned at some distance from the reactive metal interfaces, to track diffusivities using cell voltages, as presented by \textcite{Farkhondeh2017}. 

When performing restricted diffusion measurements, one should ensure that the inter-electrode length, $L_\textrm{s}$, between electrodes and duration of voltage relaxation, $t$, are such that the voltage relaxes linearly. Other indicators of composition such as conductance and spectroscopic response can also been used to track concentration gradients between two points \autocite{Harned1945,Chapman1973,Thompson1989,Stewart2008}.\\

\textbf{Galvanostatic polarisation.}
The galvanostatic polarisation, or current interrupt method, originally implemented by \textcite{Ma_1995} for polymer electrolytes, can be used to measure the transference number. This analysis is based on semi-infinite linear diffusion \autocite{Bard2001}, where the concentration gradient or OCV measured immediately after a galvanostatic pulse is directly related to the transport and thermodynamic properties of the electrolyte. By varying the polarisation time, $t_\text{i}$, and current density, $i_\mathrm{app}$, the slope, $m$, of cell potential, $V$, after the moment of current cutoff with respect to a transformed time variable $i_\mathrm{app}t_\text{i}^{1/2}$, can be related to the transference number via:
\begin{equation}
    \tp = 1-\frac{mFc_{\mre0}\sqrt{\pi D_\mre}}{4\frac{dV}{d\ln c_\mre}}.
\end{equation}
Here $F$ is the Faraday constant, $c_{\mre0}$ is the bulk salt concentration, $D_\mre$ is the diffusivity and $\frac{dV}{d\ln c_\mre}$ is the concentration dependence of the derivative of the measured voltage (the diffusion potential) with respect to the logarithm of the concentration. Note that this method requires independent concentration cell and diffusion coefficient measurements to determine $D_\mre$ and $\frac{dV}{d\ln c_\mre}$, and so care must be taken to preclude the propagation of experimental errors. While valid for non-ideal electrolytes, this relationship is derived using concentrated solution theory, assuming a binary electrolyte with electrode reactions under anion-blocking conditions. Several groups have used the galvanostatic polarisation method to measure transport properties in non-aqueous LiB electrolytes \autocite{Ehrl2017,Nyman2008,Landesfeind2019,Lundgren2015,Zugmann2011}. \textcite{Farkhondeh2017} have also fitted their four-electrode galvanostatic-pulse relaxation experiments with a transport model, to extract electrolyte transport properties.\\

\textbf{Potentiostatic polarisation.}
The potentiostatic polarisation, or steady-state direct current, method was popularised by \textcite{Bruce1987} for transference measurements in polymer electrolytes. The property measured by this method is better described as a current fraction \autocite{Galluzzo2020} or transport number, $t_+$ \autocite{Hou2020,Wang2020,Lundgren2014}, since infinite dilution is assumed in the original derivation of equation \ref{eq: BVtransference} --- implying that ionic species are non-interacting. It should be emphasised that the interpretation of parameters $t_+$ and $\tp$ is distinct. Analysis by \textcite{Bergstrom2021} has shown that potentiostatic polarisation measurements with liquid electrolytes are highly susceptible to the instability of lithium metal electrode surfaces. As such, $t_+$ cannot be used to parameterise models that accurately predict ion flux through the electrolyte phase.

The original Bruce--Vincent method applied a small constant potential bias (10 mV) and compared the initial and steady-state currents, after subtracting interfacial resistance effects, to determine the transference number via:
\begin{equation}
    \label{eq: BVtransference}    
   t_+ = \frac{I_\text{ss}(V - I_\text{i} R_\text{i})}{I_\text{i}(V - I_\text{ss} R_\text{ss})}
\end{equation}
Here, $I_\text{i}$ is the initial current measured when a bias voltage, $V$, is applied, and $i_\text{ss}$ is the steady-state current density. $R_\text{i}$ and $R_\text{ss}$ are the interfacial resistances (measured with EIS) from the electrodes, before and after polarisation. \textcite{Hou2020} have demonstrated that the propensity to underpredict concentration polarisation due to $t_+$ may be corrected for by accounting for the partial molar volume of salt. Recently,  \textcite{Balsara2015} have rederived the relationship between the steady-state current and transference number from concentrated solution theory, injecting the dependence of the diffusion coefficient and thermodynamic factor for higher salt compositions. Comparisons between galvanostatic and potentiostatic polarisation methods for transference numbers have been provided by \textcite{Zugmann2011} and \textcite{Ehrl2017}.\\

\textbf{Hittorf method.} The Hittorf method measures the change in moles of cations, compared to anions, across the cell electrolyte cavity, with the passage of a set amount of charge \autocite{Robinson2002}. This ratio is then directly related to the transference number. \textcite{Hou2020} designed a densitometric Hittorf cell with stopcocks, allowing the isolation of anodic and cathodic chambers after polarisation. Once gradients in the isolated chambers relax, the changes in density between chambers could be used to determine the concentration difference. The transference number is related to the concentration difference via:
\begin{equation}
   \tp = 1 + \frac{FV_\text{chamber}\Delta c_\mre}{Q(1-\overline{V}_\mre c_\mre)},
\end{equation}
where $V_\text{chamber}$ is the volume of the cell chamber where ion accumulation or depletion occurs, $\Delta c_\mre$ is the final change in concentration between the neutral chamber and cathodic or anodic chambers, $Q$ is the charge passed and $\overline{V}_\mre$ is the partial molar volume of salt attained from the density-molarity correlation. The overall change in composition from the Hittorf method may also by tracked conductometrically and potentiometrically \autocite{Valoen2005}.

\subsubsection{Concentration cells} \label{sec:conccell}

Concentration cells place two electrolytes with similar molecular species in chemical contact, to measure the liquid-junction potential that arises between them. This liquid-junction potential owes to the difference in chemical potential incurred by the different salt concentrations across the cell. When the OCV difference is measured with reversible lithium metal electrodes across the lithium-ion electrolyte concentration cell, the liquid-junction potential, $\phi_\textrm{e}$, is related to the thermodynamic factor, $1+\partial\ln f_{\pm}/\partial\ln c_{\textrm{e}}$, and transference, $\tp$, via the change in voltage with respect to $c_\mre$ \autocite{NewmanElectrochemical}:

\begin{equation} \label{conccell}
   \frac{\partial \phi_\textrm{e}}{\partial \ln c_\mre} = \frac{2RT}{F}\left(\tdf\right)\left(1-\tp\right).
\end{equation}

\noindent The above equation can be derived by taking equation \ref{fc2} at open-circuit (zero current) conditions.

Diffusion is controlled by placing both electrolytes in contact through a porous separator or frit with low porosity. By varying the reference and test concentrations on either side of the cell, the combination of the thermodynamic factor and the transference number that appears in equation \eqref{conccell} can be extracted. The thermodynamic factor for lithium-ion electrolytes is then isolated by an independent measurement of transference number via polarisation--relaxation experiments \autocite{Stewart2008,Valoen2005,Hou2020,Wang2020,Lundgren2014,Lundgren2015,Nyman2008,Landesfeind2019}. Concentration-cell experiments inherently average the transference number across the concentration profile through the frit. The difference in test and reference concentrations should therefore be minimised, to prevent convolution of concomitant changes in transference and thermodynamic factors \autocite{Wang2020}.

\subsubsection{Spectroscopic techniques}
Spectroscopic techniques, such as NMR spectroscopy, have been used to characterise electrolyte transport properties, largely from two directions. The first is using spectroscopy to extract entire lithium-ion concentration profiles. Often this is done with \textit{in situ} magnetic resonance imaging (MRI). Similar polarisation-relaxation experiments are performed, but MRI is used in place of voltammetric or densitometric indicators of concentration. These profiles are used to validate \autocite{Wang2021} or inversely fit relevant transport properties \autocite{Bazak2020, Klett2012,Sethurajan2015}. Raman spectroscopy and X-ray scattering has also been used to measure salt concentration profiles \autocite{Fawdon2021,Galluzzo2020}. Transport parameters obtained from inverse fitting are therefore highly dependent on the formulation of the models used to fit composition gradients.

Secondly, pulse-field gradient (PFG) NMR can be used to measure self-diffusion coefficients of nuclei in electrolytes, which relates to their equilibrium Brownian motion. By using radio-frequency pulses along with magnetic field gradients, spatial information on cation, anion, or solvent nuclei can then be encoded.
Because PFG-NMR does not distinguish between charged and neutral species, assumptions need to be made to translate self-diffusion coefficients into transport properties of charge carriers \autocite{Krachkovskiy2017}. Srinivasan's research group connected PFG-NMR results with Onsager-Stefan-Maxwell concentrated solution theory, by calculating binary interaction parameters using a Darken relation between individual diffusivities \autocite{Kim2016,Feng2017}. Discrepancies in PFG-NMR results may stem from the lack of electric field and chemical potential driving forces, which are present during battery operation. An emerging technique involves applying a large electric field to PFG-NMR, known as electrophoretic NMR. The measured instantaneous velocity between species is then used to interpret transport properties \autocite{Gouverneur2015,Timachova2019,Villaluenga2018}.

\subsubsection{Densitometry}
\label{sec:densitometry}

The electrolyte density can easily be attained by comparing the mass and volume of a particular solution. It is commonly measured with commercial densitometers, that operate based on the oscillating U-tube principle \autocite{Nyman2008,Hou2020,Wang2020,Bazak2020,Fawdon2021}. By vibrating a container filled with solution, the frequency is then linked to the mass per unit volume. 

Partial molar volumes, as discussed in section \ref{sec: PMV}, give the change in volume of a solution per mole of solvent, $\overline{V}_0$, or salt, $\overline{V}_\mre$, added. The partial molar volumes are directly related to the way density, $\rho$, changes as a function of salt concentration, $c_\mre$, \autocite{NewmanElectrochemical}:

\begin{equation} \begin{array}{l}
\dst  \overline{V}_\mre = \frac{M_\mre - \frac{d\rho}{dc\mre} }{\rho - c_\mre \frac{d \rho}{dc_\mre}}, \quad 
 \overline{V}_0 = \frac{M_0}{\rho - c_\mre \frac{d \rho}{dc_\mre}},
    \label{eq:PMV}
\end{array} \end{equation}

\noindent where $M_0, M_\mre$ are the respective molar masses of solvent and salt.

\section{\label{bass} Parameter database}

As part of this review, we provide a free and open database of battery, and DFN model, parameters called LiionDB, available at {\tt www.liiondb.com} \autocite{LiionDB}. Open battery data repositories like this have aided researchers across the energy storage spectrum. For example, the Battery Archive is a recently published open-source life cycle repository \autocite{BatteryArchive}. \textcite{dosReis2021} have also summarised public sources of battery cycling aging data for different chemistries and operating conditions.  Other tools, such as the Battery Explorer by the Materials Project, provide a database of material properties that can accelerate compound discoveries \autocite{Jain2013}. Other text-mining approaches have been applied to parse a large quantity of published literature, to extract relevant material properties \autocite{Huang2020,ElBousiydy2021}.

The complexity in DFN property relations precluded auto-parsing of literature parameters via data-driven techniques in this review. Hence, parameters were abstracted individually by reproducing values or correlations where provided, or extracted manually from published plots. 

\begin{figure}
  \centering
  \includegraphics[width=0.8\textwidth]{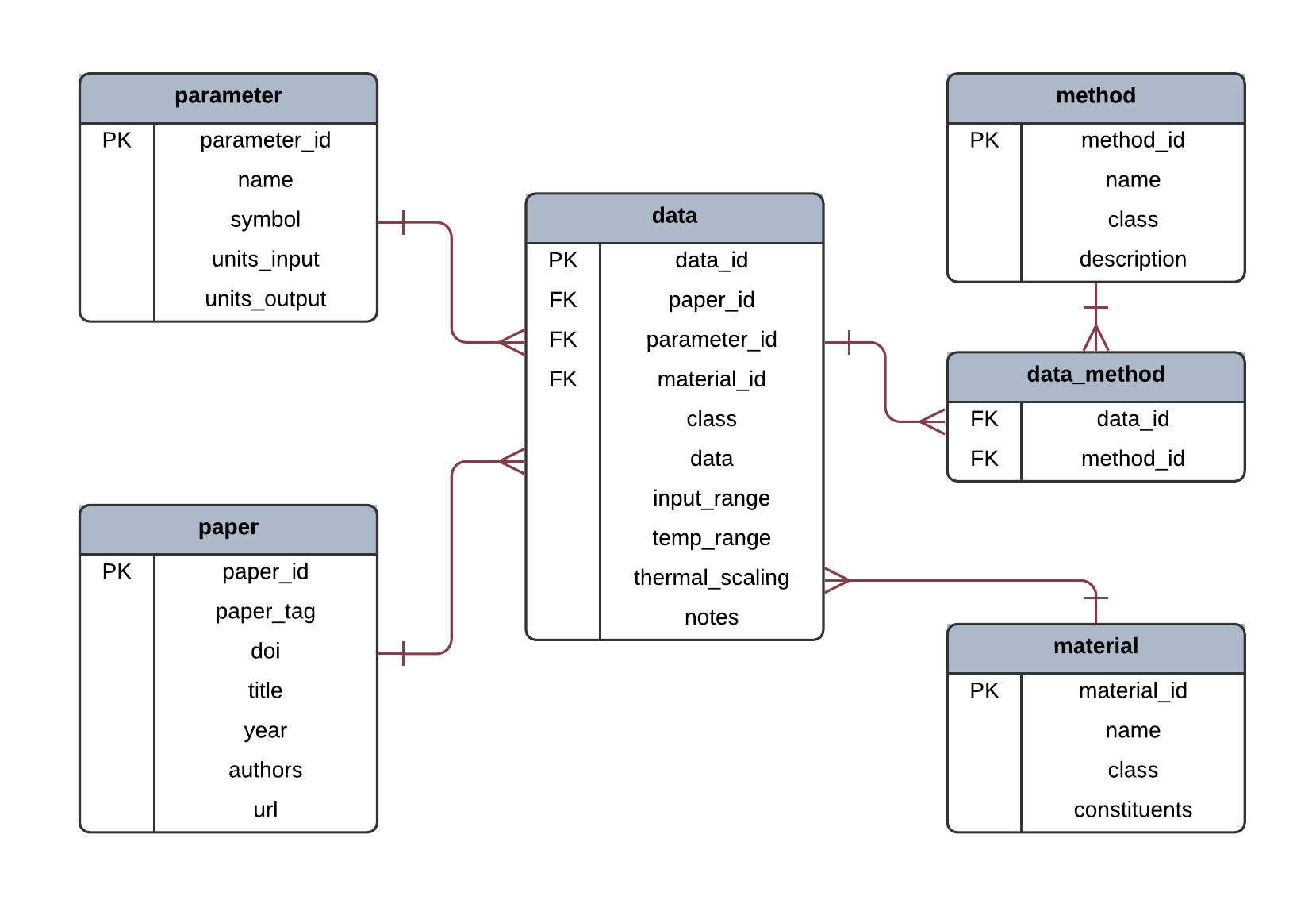}  
\caption{LiionDB parameter database structure \autocite{LiionDB}. For a set of data extracted from literature, there can be a DFN parameter, paper, material and methods associated with it. Each table is related to each other through primary keys (PK) and foreign keys (FK).}
\label{fig:ERD}
\end{figure}

The database parameters reported alongside this manuscript were limited to literature sources for lithium-ion cell chemistries, with a focus on relevance to DFN-based battery models. They are presented in a searchable PostgreSQL format, along with a GUI web application (at {\tt www.liiondb.com}) \autocite{LiionDB}. A schematic of the database structure is depicted in figure \ref{fig:ERD}. At the time of publishing, there are 1000+ parameter data sets in scalar and functional forms across the materials and papers included. Each data set is also associated with the methods broadly employed during its direct or inverse measurement. The applicable composition range, temperature and thermal scaling methods are also stated, where available. Material sub-components have also been individually included, to maintain search flexibility. 

The authors caution against parameterising battery models without regard for where they came from. Best practice guidelines for battery modelling should be followed \autocite{ModelChecklist}. This database acts as a directory for a range of parameter values reported in the literature, allowing modellers access to the context of how the physical parameters were measured. As covered in sections \ref{section:geom mft}, \ref{section:electrode mft} and \ref{section:elyte mft}, many properties are method-specific and compatibility with model formulation should be confirmed, to ensure physical consistency.

\paragraph{Contributing to the database} We encourage the community to contribute towards LiionDB. On the website {\tt www.liiondb.com} \autocite{LiionDB}, we have provided information on how authors can submit their data for inclusion. This will not only serve the community in general, by making the database more comprehensive, but also offers contributors better visibility for their work, leading to more citations and interaction with other research groups around the globe.

\section{\label{conc}Discussion and conclusions}
The DFN modelling framework has become a ubiquitous tool for carrying out cell-scale simulations of LiBs. Despite its widespread use, challenges remain around how to fully and accurately determine the scalar and functional parameters that are required for its use. Whilst a great deal of characterisation has been reported in the literature, there was no central repository where independent measurements are collated and can be compared. The database that accompanies this review (available at {\tt www.liiondb.com}) fills this need and we hope that it will serve the battery modelling community, by becoming the main resource for finding DFN model parameter values.

The second purpose of this review is to inform modellers and experimentalists alike about some of the often unstated assumptions implicit in the methods used to infer parameter values from raw experimental data. The aim here is to prevent researchers from using parameter values which were extracted under assumptions inconsistent with the operating conditions in which they aim to use their model. This is critical, to ensure that models have genuinely robust predictive power. For many parameter values, the consistency of assumptions is clear, for example, in order to have truly accurate parameter values for the thickness and porosity of the separator layer, one should measure these properties under the mechanically loaded conditions that would be encountered in a working cell. For other parameter values, for example, the solid-state diffusivity as inferred by electrochemical impedance spectroscopy, there are typically numerous (and more subtle) assumptions at play, e.g., whether insertion materials are treated as spherical or planar with semi-infinite geometry. The parameterisation challenge is further complicated by the issue of identifiability: A nondimensionalisation of the equations can be carried out to reveal which parameters can be learned from a given experiment and which parameters are required to parameterise a model. We hope that this review serves to make these assumptions clear, thereby empowering researchers to make informed choices about which parameter values are most appropriate for their particular situation. 

A compounding issue that applies to many parameters is the propagation of errors in the parameterisation process. Consider again the solid-state diffusion coefficient. The model used to convert the raw experimental data to a diffusion coefficient requires the input of geometrical properties, e.g., the electrochemically active surface area of the electrode particles. This quantity is almost always not known exactly and instead has to be estimated by another experimental procedure. Thus, we must be mindful of how these errors can potentially propagate.

Although this review has focused primarily on the DFN model, there are many applications in which reduced models give sufficient predictive power. For example, a cell phone battery discharging at a modest C-rate is well-described by a so-called `single particle model'. In such cases, it is prudent to utilise these reduced models, not only in the interests of computational efficiency, but also because they typically contain fewer parameters and are therefore simpler to parameterise. 

Despite all these difficulties, the DFN framework is an extremely powerful tool and has been used to great effect in helping to accelerate LiB development.

\section{\label{Acknowledgements}Acknowledgements}

The authors thank the Faraday Institution (https://faraday.ac.uk/; EP/S003053/1), grant number FIRG003, for funding. 


\printbibliography

\end{document}